\documentclass{elsart}

\usepackage{graphicx}

\usepackage{amssymb}
\usepackage{amsmath}
\def\be{\begin{eqnarray}}
\def\ee{\end{eqnarray}}
\def\ba{\begin{array}}
\def\ea{\end{array}}

\def\g{\noindent}
\def\ds{\displaystyle}

\def\mm#1{\mbox{\bf #1}}

\def\erf{\mbox{erf\hskip1pt}}

\begin{document}

\begin{frontmatter}



\title{Coupling to the continuous spectrum and HFB approximation}


\author[label1]{K. Bennaceur,}
\ead{k.bennaceur@ipnl.in2p3.fr}
\author[label2]{J.F. Berger},
\author[label2]{B. Ducomet}
\address[label1]{IPN Lyon, CNRS IN2P2 Universit\'e Claude Bernard Lyon I, \\
       43 Bd du 11 Novembre 1918, \\
       69622 Villeurbanne Cedex, France}
\address[label2]{CEA-D\'epartement de Physique Th\'eorique et Appliqu\'ee, \\
       BP 12, 91680 Bruy\`eres-le-Chatel, France}

\begin{abstract}
We propose a new method to solve the Hartree-Fock-Bogoliubov equations
for weakly bound nuclei whose purpose is to improve the treatment of
the continuum when a finite range two-body interaction is used.
We replace the traditional expansion on a discrete harmonic oscillator
basis by a mixed eigenfunction expansion associated with a potential
that explicitely includes a continuous spectrum. We overcome the problem
of continuous spectrum discretization by using a resonance expansion.
\end{abstract}

\end{frontmatter}

\section{Introduction}
\label{seca}

In many problems of present-day nuclear physics (structure of the ground
state of nuclei close to drip lines, description of weakly bound or unbound
excited states, decay of isomeric states), one has to explicitly take into
account the continuum of scattering states in addition to the discrete
spectrum of bound states. A convenient approach to these problems is provided
by the mean-field theory in which pairing correlations are included in a
self-consistent way through the Hartree-Fock-Bogoliubov (HFB)
formalism~\cite{dnwbcd,dg}. In this method a set of quasi-particle
excitations together with their vacuum is obtained, which forms the basis
for a description of ground state properties and spectra. In principle,
quasi-particle states, when expressed in terms of the single-particle
eigenstates of a one-body hamiltonian, are always superpositions of states
belonging to both the discrete and continuous
spectra~\cite{dft,bul}.
In ordinary stable nuclei, the continuum part of quasi-particle
states is usually neglected, as ground state properties and low-energy
excitations mainly depend on the discrete quasi-particle
states representing occupied and negative energy unoccupied single-particle
orbitals. So, quasi-particles are often obtained using
expansions on a discrete basis of orthonormal square-integrable functions.

A very different situation is encountered in weakly bound nuclei, where
the gap between the last occupied single-particle level and the continuum
of positive energy states may become smaller than a few MeV. In this case,
the residual pairing interaction is able to induce a significant coupling
between the discrete and continuum single-particle states and the
contribution of the continuum to quasi-particle states cannot be neglected.

In such a situation, new numerical methods for solving the HFB equations
have to be implemented, which more or less implicitly, amount to introduce
a suitable discretization of the continuum. With these methods, not only
the contribution of the continuum to quasi-particle states can be
obtained, but also a description of quasi-particles belonging to the
continuous spectrum, including resonances. Let us note that such a
treatment is necessary to reproduce the unusual properties of weakly
bound nuclei such as halos and neutron skins.

Several techniques have been proposed in the past along these lines: lattice
calculations~\cite{bonche}, Basis-Spline Galerkin lattice~\cite{ober},
methods using the local-scale point transformation of the spherical
harmonic oscillator wave functions~\cite{stoitsov}, which are adapted to
calculations involving zero-range two-body interactions such as the Skyrme
forces. Another approach, based on the Kamimura-Gauss basis~\cite{kg} has
been reported recently~\cite{nakada}. It appears to be a promising
technique for both zero range and finite range forces. Recently, a method
has been proposed where the HFB equations are solved in a spherical
box with exact boundary conditions for scattering
states~\cite{gsgl,ggs}. In these methods, the HFB equations are
usually expressed in space coordinates, which is especially convenient
when a zero-range nucleon-nucleon interaction is used, and
interesting results have been obtained in this context~\cite{dft,fayans,bdp}.

In the case of the finite range Gogny interaction~\cite{dg}, the HFB
equations have always been solved by expanding the quasi-particle
states on a finite discrete basis of orthonormal functions. In view of
computational convenience, the latter are usually chosen as the
eigenfunctions of an harmonic oscillator (HO), whose geometry and
symmetries are adapted to the nuclear states to be studied.
The HFB equations are then expressed in matrix form, and they are solved
by using an iterative procedure. Namely, at iteration $n$,
the set of individual wave functions $\{\psi_i^{(n)}\}$ is expanded on a
finite basis of generalized eigenstates of a reference hamiltonian
$H^{(n)}$, generally taken as $T+V^{(n)}_{HO}$, where $V^{(n)}_{HO}$ is
an harmonic oscillator potential:
\begin{equation}
  \psi_i^{(n)}=\sum_{p=1}^{N_{max}}C_{i,p}^{(n)} \varphi_{p}^{(n)}.
  \label{bog15}
\end{equation}

\g The $\varphi_{p}^{(n)}$ are explicitly known and the unknown coefficients
$C_{i,p}^{(n)}$ are determined using matrix diagonalization techniques.

This simple choice has been found to be appropriate for computing the
properties of well bound nuclei, the convergence of the iterative
procedure being obtained with a reasonable number of
iterations~\cite{dg,gg}.
In practice, the sum in (\ref{bog15}) is truncated to a number of states
$N_{max}$ corresponding to a number of HO shells ranging
from 6-8 in light nuclei up to 16-20 in heavy and superheavy nuclei.

On the other hand, in exotic nuclei close to drip-lines, weakly bound
single-particle states acquire a spatial extension so large that
the use of expansions on HO wave functions such as~(\ref{bog15}) requires
prohibitively large values of $N_{max}$.
In this case, as mentioned above, an alternative method of solving the HFB
equations has to be found.

A natural way to avoid this difficulty is to change the reference potential
from which the basis is built. Let us recall the well-known
``decomposition of unity" on a complete basis of generalized eigenstates,
associated with a reference hamiltonian, a sum of kinetic
and potential terms $H=T +V$. We assume that  $V$ is short-ranged and
negative for small $r$.
This decomposition formula can be written formally as
\begin{equation}
{\mathbf 1}=\sum_{n=1}^N | \varphi_n><\varphi_n| +\int
| \varphi(\cdot,k)><\varphi(\cdot,k)|\ dk,
 \label{comp}
\end{equation}
where the first (resp. second) contribution corresponds to the discrete
(resp. continuous) spectrum of $H$.

Numerical simulations~\cite{dugro,dugro2} show that the continuous
contribution in (\ref{comp}) can be neglected in the
expansion of single particle states provided the gap between the energy
of the last occupied state $|E_N|$ and the continuous threshold
is large enough, which is the case in stable nuclei.

As mentioned before, for nuclei near instability, the gap is small and
the continuous contribution in (\ref{comp}) has to be taken into account.

A new difficulty then appears: a numerical treatment based on
equation~(\ref{comp}) requires to introduce both a discretization and a
cut-off in the integral involving continuous states. This leads to a finite
expansion upon discrete basis states which is not necessarly
less expensive than the one in (\ref{bog15}), unless one can identify a few
states carrying a predominant
contribution of the continuous part in (\ref{comp}). As shown in a simplified
model~\cite{ducpair}, such a discretization would probably lead to
the same numerical difficulties as above, as the necessary truncation of
the integral in(~\ref{comp}) is not likely to be less expensive than a sum
with large $N_{max}$ in~(\ref{bog15}).

As a matter of fact one knows that the so-called metastable or resonant
states corresponding to complex singularities
of the resolvent give important contributions to the density of states.
These states decay exponentially in time, preventing them to
be considered as part of the spectrum, although they correspond to
long-living nuclear
configurations. Moreover it has been soon recognized (see~\cite{kukulin}
and references therein) that, despite their pathological asymptotic behaviour,
these unstable states could be used as generalized eigenfunctions that can
be used to describe resonance phenomena in nuclear collisions. In this
framework, a given state  is expanded on a set of functions that
includes resonant functions on the same footing as bound state eigenfunctions.

The drawbacks of such a procedure are well known: as the resulting
problem looses its self-adjointness, the complex spectrum has to be
reinterpreted together with the associated wave functions.
However such expansions have been frequently used to describe various physical
situations (potential and obstacle scattering, nuclear reactions...~\cite{be}).
Recently, this approach has also
been used to treat the problem of multiconfiguration mixing
within the nuclear shell model~\cite{nicolas}.

 Our purpose is to show that this old idea may be applied in a simple way
to self-consistent computations when loosely bound or metastable states
are expected, which is precisely the case in HFB computations for heavy
nuclei, especially those which are close to drip-lines.

This paper is organized as follows. In Sec. \ref{secb}, we consider a model
where the Schr\"odinger equation is solved in a bounded domain,
and we apply it to a simplified one-body problem. In Sec. \ref{secc} some
numerical experiments are presented, while Sec. \ref{secd} briefly
presents our conclusions.

\section{The problem in a bounded domain}
\label{secb}

Classically, in order to derive generalized eigenfunction expansions such
as~(\ref{comp}), one first computes the Green function associated to
the problem and one uses complex analysis to continue this Green function
and study its singularities. Then, by applying the Cauchy residue
theorem, one obtains (\ref{comp}), where the sum corresponds to the discrete
spectrum, and the integral takes into account the continuous part of the
spectrum (see Appendix A).

A natural way to avoid the delicate discretization of the continuous
contribution in (\ref{comp}) is to restrict the initial problem
to a bounded region $\Omega$, where one expects that most of the physics
takes place, and to impose ``transparent'' boundary conditions reflecting the
properties of the exterior (unbounded) domain.
In fact, this last point deserves a special comment. It is clear that
the kind of artificial conditions we impose on the boundary must be of
particular nature in order to mimic the behaviour of the ``exterior world",
in a transparent way: the waves
produced in the interior region must not be disturbed (reflected) by the
boundary. For example, a standard Dirichlet
or Von Neumann boundary
conditions would lead to a problem essentially different
from the one we want to solve and would produce
results without clear connection with the solution of our original problem.
One can check easily in particular that Dirichlet
condition would lead to nothing but a Fourier series expansion, which makes
the obtained solution strongly domain-dependent.

One can show that this kind of ``transparent" boundary conditions, which has
been developped in the acoustic domain and more recently in the
Schr\"odinger context (see~\cite{fj} and references therein) leads to a
complex condition producing a non-selfadjoint problem: this is the price to
pay for ``replacing the exterior by the boundary".

By restricting the domain, one can hope to get improved asymptotic properties
for the modified Green functions, allowing one to bypass the
integral along the physical cuts in (\ref{comp}). As a consequence, complex
isolated singularities appear outside the imaginary axis, coming
from the artificial complex conditions at the boundary of $\Omega$.

To our knowledge, this idea goes back to Kapur and Peierls~\cite{kapu}
who first introduced in the thirties non self-adjoint flux
conditions associated to Schr\"odinger equations, a method which has been widely used in the context of nuclear reactions~\cite{kukulin}.

\subsection{The model}

In the particular situation of the square well potential, the above method
can be worked out in a very simple way.
The idea is to replace the harmonic oscillator by a new reference potential
$V$ such that the hamiltonian $H=T+V$ has a continuous
spectrum, in order to describe more accurately the loosely bound states and,
if necessary, metastable states.

As we expect that the physics takes place in the region $\Omega$ where the
potential is non-zero, we restrict the dynamics to $\Omega$
and we impose a boundary condition at the boundary which mimics the external
scattering behaviour.

To be specific, let us consider the Schr\"odinger equation for a particle
in a compactly supported spherically symmetric potential $V$:
\begin{equation}
{\displaystyle \left( -\frac{d^2}{dr^2}+V(r)\right)\varphi_k(r)
=k^2\varphi_k(r),\ \ \mbox{for}\,\ r\,\in\,\Omega=(0,r_0).}
\label{schrodinger_radial}
\end{equation}
Where we have considered $\hbar^2/2m=1$ and an orbital momentum $\ell=0$.
As a new reference potential $V$, we take the square well defined as:
\begin{equation}
V(r)=
\left|  \begin{array}{ll}
{\displaystyle -V_0,\ \mbox{for}\ r<r_0     },\\[3mm]
{\displaystyle 0         ,\ \mbox{otherwise}},
 \end{array}
\right.
\label{pui}
\end{equation}
where $V_0>0$ and $r_0>0$.

We chose the square well because the associated eigenfunctions are
rather elementary (Bessel functions). Moreover, in order to simplify the
arguments we restrict the analysis to $s$-waves, in which case the
eigenfunctions are simply sine and cosine functions.

As an extra boundary condition for $r=r_0$, we choose the free radiation
condition:
\begin{equation}
{\displaystyle \left(\frac{d}{dr}-i k\right)\varphi_k(r)=0,
\ \ \mbox{for}\ r=r_0.}
\label{bord}
\end{equation}
 From the above discussion about ``transparent" boundary conditions, we
stress that the non-selfadjoint boundary condition (\ref{bord})
amounts to fix an imaginary flux condition reflecting exactly the physical
radiation behaviour at infinity for the free Schr\"odinger equation.
We feel that this choice corresponds to our purpose to produce the smallest
perturbation near the artificial boundary $r=r_0$.

The problem (\ref{schrodinger_radial})-(\ref{bord}) is easily solved by
computing the corresponding Green function (see Appendix A), giving
simple explicit expressions for the associated eigenfunctions and the
following expansion for any function $f\in L^2(0,r_0)$:
\begin{equation}
{\displaystyle
f(r')=
\sum_n
\frac{k_n}{i + k_nr_0}
\left(\int_0^{r_0} \sin p_nr    f(r)\ dr \right) \sin p_nr',}
\label{expan}
\end{equation}
where the $k_n$ are the solutions both, real and complex, of the equation :
\begin{equation}
p_n\cot p_nr_0=i k_n,
\label{treq}
\end{equation}
with $p_n=(k_n^2+V_0)^{1/2}$.
For real roots we recover bound and virtual state eigenfunctions, while for
non real ones we get resonant and anti-resonant eigenfunctions.

\subsection{Application to HFB problems}

Let us consider the familiar Hartree-Fock-Bogoliubov equations:
\begin{equation}
\left(\begin{matrix}
  T+{\mathcal U} & {\mathcal V} \\
  {\mathcal V} & -T-{\mathcal U}
\end{matrix}\right)
\left(\begin{matrix}
  \psi_1\\ \psi_2\
\end{matrix}\right)
=\left(
\begin{matrix}
  E+\lambda & 0\\
  0 & E-\lambda
\end{matrix}\right)
\left(\begin{matrix}
  \psi_1\\ \psi_2
\end{matrix}\right),
\label{eqhfb}
\end{equation}
where the particle-hole and particle-particle potentials ${\mathcal U}$
and ${\mathcal V}$ depend on $\psi_{1}$ and $\psi_{2}$.

In order to solve this non linear eigenvalue problem we expand the unknown
wave functions $\psi_1$ and $\psi_2$ on our reference basis,
suitably truncated up to a rank $N$, depending on the desired accuracy:
\begin{equation}
\psi_1=\sum_{n=0}^N a_n\varphi_n,\mbox{\ and\ \ }
\psi_2=\sum_{n=0}^N b_n\varphi_n,
\label{fifi}
\end{equation}
where
\begin{equation}
\varphi_n(r)=\sin p_nr
\label{eqset}
\end{equation}
and the coefficients $p_n$ are obtained by solving the transcendental
equation (\ref{treq}).

\smallskip

Replacing the functions $\psi_1(r)$ and $\psi_2(r)$ by there expansions,
multiplying on the left by $\varphi_m^*(r)$
and integrating, one gets the following expressions for (\ref{eqhfb}):
\begin{equation}
\left\{\begin{matrix}
\ds\sum_{n=0}^N
 (\mathbf T_{mn} + \mathbf U_{mn}) a_n+\mathbf V_{mn} b_n\,=\,(E+\lambda)
\sum_{n=0}^N\mathbf R_{mn}a_m, \\
\noalign{\vskip 0.2em}
\ds\sum_{n=0}^N
 \mathbf V_{mn} a_n
-(\mathbf T_{mn} + \mathbf U_{mn}) b_n\,=\,(E-\lambda)
\sum_{n=0}^N\mathbf R_{mn}b_m.
\end{matrix}\right.
\label{eqvpg}
\end{equation}
or in matrix form:
\begin{equation}
\left(\begin{matrix}
  \mathbf T+\mathbf U & \mathbf V \\
  \mathbf V & -\mathbf T-\mathbf U
\end{matrix}\right)
\Psi=
\left(\begin{matrix}
  (E+\lambda)\mathbf R & 0 \\
  0 & (E-\lambda)\mathbf R
\end{matrix}\right)
\Psi
 \label{eqvpga}
\end{equation}
where {\bf T}, {\bf U}, {\bf V} and {\bf R} are complex hermitian matrices
and $\Psi$ is the column vector of components
$c_i\equiv(a_1,...,a_N,b_1,...,b_N)$.

The equation (\ref{eqvpga}) is a generalized eigenvalue problem, from which
the coefficients of the expansion (\ref{eqset}) together with
the eigenvalues $E$ can be obtained.

If we solve the problem using only the functions $\varphi_n$ (corresponding
to $\ell=0$), these elements are rather simple and in many cases
they can be computed analytically.

Some care must be taken in the evaluation of the kinetic energy ${\mathbf T}$.
As discussed before, the expansion we use is valid only for
$r\in (0,r_0)$ and expansion (\ref{fifi}) concerns only the part of the
functions $\psi_{1,2}(r)$ defined in this interval. Consequently
equations~(\ref{fifi}) have to be replaced by:
\begin{equation}
Y(r_0-r)\psi_1(r)=\sum_n a_n\varphi_n(r),\ \mbox{and}\ \ Y(r_0-r)\psi_2(r)=
  \sum_n b_n\varphi_n(r),
\end{equation}
where $Y(r_0-r)$ is the Heaviside function.
Taking the second derivative of this function, one gets:
\begin{equation}
\frac{d^2}{dr^2}\Bigl[Y(r_0-r)\psi_i(r)\Bigr]=
Y(r_0-r)\frac{d^2\psi_i(r)}{dr^2}-\delta(r_0-r)
\left.\frac{d\psi_i(r)}{dr}\right|_{r=r_0}
\end{equation}
As a consequence, the matrix elements of the kinetic energy $T$ are:
\begin{align}
  \begin{split}
    \ds \mathbf T_{mn} &= \ds -\frac{\hbar^2}{2m}  \int_0^{r_0}
      \varphi_m^*(r)\biggl[\Delta-\delta(r_0-r)\frac{d}{dr}\biggr]
      \varphi_n(r)\ dr \\
  \noalign{\vskip 3pt}
                     &= \ds \frac{\hbar^2 k_n^2}{2m} \mathbf R_{mn}
      +\frac{\hbar^2 k_n^2}{2m}\sin(k_mr_0)\cos(k_nr_0)
  \end{split}
  \label{eqemt}
\end{align}
They are computed in the same way as those of {\bf R} (see relation
(\ref{eqvpg})).
Let us note that similar formulas have to be employed if the potentials
${\mathcal U}$ and ${\mathcal V}$ contain velocity-dependent
contributions for example in the spin-orbit components.

In the case of a nucleon-nucleon force of simple form (Skyrme or Gogny
forces), the matrices ${\mathcal U}$ and ${\mathcal V}$ can be calculated
analytically. Let us note that since {\bf T} and {\bf R} are hermitian
matrices, the reality of the eigenvalues in equation
(\ref{eqvpg}) is insured. Finally the Coulomb potential can also be
calculated analytically.

\subsection{Canonical states}

In order to interpret the HFB results, let us introduce the normal reduced
density $\rho(r,r')$:

\begin{equation}
rr'\rho(r,r')=\sum_{k} \psi_2^{(k)}(r)\ {\psi_2^{(k)}}^*(r'),
\label{norden}
\end{equation}
where $\psi_2^{(k)}$ denotes the lower component of the $k^{th}$ solution
of equation (\ref{eqvpga}).
With a similar notation, the abnormal reduced density $\tilde{\rho}(r,r')$ is
given by:
\begin{equation}
rr'\tilde{\rho}(r,r')=-\sum_{k} \psi_2^{(k)}(r)\ {\psi_1^{(k)}}^*(r'),
\label{anorden}
\end{equation}
Using expansion (\ref{fifi}), these relations become:
\begin{equation}
rr'\rho(r,r')=\sum_{k}\sum_{i,j} b_i^{(k)} {b_j^{(k)}}^* \varphi_i(r)
\ \varphi_j^*(r'),
\label{nrden}
\end{equation}
and:
\begin{equation}
rr'\tilde{\rho}(r,r')=-\sum_{k}\sum_{i,j} b_i^{(k)} {a_j^{(k)}}^* \varphi_i(r)
\ \varphi_j^*(r'),
\label{anrden}
\end{equation}
for $0\leq r,r'\leq r_0$.

The interpretation of these quantities is clear (see for
example~\cite{dnwbcd}). The normal density is localized as soon as the
chemical parameter $\lambda$ is negative. The canonical basis
$\{ \overline{\psi}_i\}$ is defined as the set of eigenvectors of the
normal density:
\begin{equation}
\int dr'rr'\rho(r,r')\ \overline{\psi}_i(r')=v_i^2 \overline{\psi}_i(r),
\label{bascan}
\end{equation}
where the eigenvalues $v_i^2$ are the occupation probabilities of the
corresponding canonical states. It is straightforward to show that
if the $\overline{\psi}_i$ are expanded in the same way as the quasi-particle
wave functions, this equation corresponds to:
\begin{equation}
{\mathbf\rho}\ {\mathbf R}\ \overline{\Psi}_i=v_i^2 \overline{\Psi}_i,
\label{baca}
\end{equation}
where $\overline{\Psi}_i$ is the vector built with the coefficients
of the expansion of $\overline{\psi}_i(r)$ and $\rho$ is the matrix
defined by:
\begin{equation}
\rho_{ij}=\sum_k b_i^{(k)} {b_j^{(k)}}^*.
\end{equation}
It is important to stress that $\rho_{ij}$ is not properly speaking
a representation of the normal density since the set of functions
$\{\varphi_i\}$ does not form a basis.

Finally, we define the energies of the canonical states as the
diagonal matrix element of the Hartree-Fock Hamiltonian in
the basis of the canonical states~\cite{dnwbcd}:
\begin{equation}
\epsilon_i = \langle \overline{\Psi_i}|H|\overline{\Psi_i}\rangle
\end{equation}
This term can be easily written once the coefficients of the expansion of
the canonical states are known.

\section{Numerical simulations}
\label{secc}

To illustrate this method we have chosen to solve a simplified problem of HFB
type with spherical symmetry. We consider the following equations:

\begin{equation}
\left(\begin{matrix}
  T+{\mathcal U} & {\mathcal V} \\
 {\mathcal V} & -T-{\mathcal U}
\end{matrix}\right)
\left(\begin{matrix}
  \psi_1\\ \psi_2
\end{matrix}\right)=
\left(\begin{matrix}
  E+\lambda & 0\\ 0 &E-\lambda
\end{matrix}\right)
\left(\begin{matrix}
  \psi_1\\ \psi_2
\end{matrix}\right)
\label{neqa}
\end{equation}
where $T$ is the kinetic energy operator, ${\mathcal U}$ and ${\mathcal V}$
are respectively the particle-hole and particle-particle
mean-fields. We do not take into account the possible effects of a Coulomb
field or of a spin-orbit potential, although
this would not lead to any difficulties in the treatment.

\subsection{Hartree-Fock case (no pairing)}

\begin{table}[htbp]
\begin{center}
{\small
\[\begin{array}{|c|r|r|}
\hline
\hline
\multicolumn{1}{|c}{\ \ \ell\ \ } & \multicolumn{1}{|c|}{E}
                                  & \multicolumn{1}{|c|}{\Gamma} \\
\hline
 0 &\ -4.571\,183 &           0 \\
 0 & -0.884\,281 &           0 \\
 0 &  2.252\,381 &\  0.000\,118 \\
 0 &  4.500\,948 &  0.247\,951 \\
 0 &  6.008\,281 &  2.516\,116 \\
 0 &  7.587\,937 &  6.266\,307 \\
\hline
 1 & -2.619\,884 &           0 \\
 1 &  0.807\,635 &      \sim 0 \\
 1 &  3.577\,387 &  0.013\,017 \\
 1 &  5.312\,240 &  1.063\,163 \\
 1 &  6.845\,729 &  4.224\,511 \\
 1 &  8.427\,005 &  8.441\,884 \\
\hline
\hline
\end{array}
\hskip 3em
\begin{array}{|c|r|r|}
\hline
\hline
\multicolumn{1}{|c}{\ \ \ell\ \ } & \multicolumn{1}{|c|}{E}
                                  & \multicolumn{1}{|c|}{\Gamma} \\
\hline
 2 &\ -0.759\,532 &           0 \\
 2 &  2.384\,152 &\  0.000\,083 \\
 2 &  4.659\,669 &  0.217\,375 \\
 2 &  6.163\,324 &  2.408\,615 \\
\hline
 3 &  1.009\,032 &      \sim 0 \\
 3 &  3.810\,922 &  0.008\,196 \\
 3 &  5.581\,406 &  0.932\,281 \\
 3 &  7.091\,723 &  4.003\,469 \\
\hline
 4 &  2.676\,525 &  0.000\,028 \\
 4 &  5.025\,176 &  0.148\,085 \\
 4 &  6.522\,260 &  2.137\,256 \\
 4 &  8.059\,380 &  5.759\,133 \\
\hline
\hline
\end{array}\]
}
\caption{Discrete levels associated with the potential defined
  by (\ref{neqpara}). The energies and widths are determined by
  looking at the position of the poles of the S matrix in the
  complex energy plane.
  \label{taba}
}
\end{center}
\end{table}
\begin{table}[htbp]
\begin{center}
\[\begin{array}{|r|r|r|r|r|}
\hline
\hline
 \multicolumn{1}{|c|}{\ell=0} &
 \multicolumn{1}{|c|}{\ell=1} &
 \multicolumn{1}{|c|}{\ell=2} &
 \multicolumn{1}{|c|}{\ell=3} &
 \multicolumn{1}{|c|}{\ell=4} \\
\hline
\mm{-4.571\,182} & \mm{-2.605\,981} & \mm{-0.759\,533} &      0.186\,392 &      0.229\,617 \\
\noalign{\vskip -2pt}
\mm{-0.884\,280} &       0.130\,394 &       0.153\,058 &      0.535\,440 &      0.585\,189 \\
\noalign{\vskip -2pt}
      0.118\,909 &       0.471\,894 &       0.497\,537 & \mm{1.009\,141} &      1.120\,116 \\
\noalign{\vskip -2pt}
      0.458\,980 &  \mm{0.836\,844} &       1.023\,303 &      1.065\,186 &      1.800\,000 \\
\noalign{\vskip -2pt}
      0.980\,599 &       0.995\,157 &       1.692\,437 &      1.738\,986 &      2.592\,748 \\
\noalign{\vskip -2pt}
      1.644\,922 &       1.660\,949 &  \mm{2.384\,031} &      2.524\,324 & \mm{2.676\,568} \\
\noalign{\vskip -2pt}
 \mm{2.252\,302} &       2.436\,584 &       2.472\,184 &      3.387\,049 &      3.466\,495 \\
\noalign{\vskip -2pt}
      2.418\,722 &       3.283\,319 &       3.327\,296 & \mm{3.811\,796} &      4.374\,556 \\
\noalign{\vskip -2pt}
      3.265\,180 &  \mm{3.617\,422} &       4.193\,201 &      4.290\,813 & \mm{4.993\,930} \\
\noalign{\vskip -2pt}
      4.106\,893 &       4.170\,489 &  \mm{4.675\,124} &      5.120\,114 &      5.392\,091 \\
\noalign{\vskip -2pt}
 \mm{4.546\,091} &       4.968\,645 &       5.215\,890 & \mm{5.742\,917} &      6.207\,120 \\
\noalign{\vskip -2pt}
      5.140\,059 &  \mm{5.605\,141} &       6.024\,047 &      6.469\,599 &      7.002\,387 \\
\noalign{\vskip -2pt}
\ \ \ \ 5.946\,712 & \ \ \ \ 6.379\,345 & \ \ \ \ 6.841\,667 &\ \ \ \ 7.342\,215 &\ \ \ \ 7.878\,341 \\
\noalign{\vskip -2pt}
      6.773\,633 &       7.256\,568 &       7.743\,484 &      8.286\,709 &      8.849\,874 \\
\noalign{\vskip -2pt}
      7.684\,636 &       8.230\,319 &       8.731\,860 &      9.300\,691 &      9.897\,122 \\
\noalign{\vskip -2pt}
      8.679\,769 &       9.252\,138 &       9.793\,386 &                 &                 \\
\noalign{\vskip -2pt}
      9.746\,902 &                  &                  &                 &                 \\
\hline
\hline
\end{array}\]
\caption{Eigenvalues obtained by diagonalizing the matrix equation
   (\ref{eqhf}). The boldface numbers correspond to the energies of the
   bound states and of the scattering states strongly localized inside
   the potential.
   \label{tabb}
}
\end{center}
\end{table}

We first consider the case without pairing, {\it i.e.}
${\mathcal V}=0$. In this case, the Lagrange multiplier $\lambda$ is
not needed any more and we can omit it in the equations. We want to
check the ability of our method to construct the correct bound
and resonant states associated with a given potential. To this aim, we
consider the gaussian potential used in~\cite{sofianos}:
\begin{equation}
{\mathcal U}(r)=5e^{-0.25(r-3.5)^2}-8e^{-0.2r^2}
\label{neqpara}
\end{equation}
This potential is given in MeV and the radial coordinate in fm in units
where $\hbar^2/2m=1/2$ MeV.fm$^2$.
This potential has a rich spectrum in all partial waves and the discretes
states have been calculated with high accuracy in~\cite{sofianos}.
The energies and widths of the discrete states up to 10 MeV are reported
in table~\ref{taba}.

To build the set of functions (\ref{eqset}) we have used a rectangular well
of depth 30 MeV and radius 12 fm. By solving equation (\ref{treq})
we have found 30 bound states and 29 virtual states. We complete
this set with the lowest resonances and companion anti-resonances.
The corresponding functions are not orthogonal and we cannot build
the matrix ${\mathbf R}$ using all of them, otherwise the
generalized eigenvalue problem (\ref{eqvpga}) would be -- at least numerically -- singular. In order to reject some functions we use the
following procedure:
we build the full  matrix ${\mathbf R}$, using all functions and try
to invert it with a simple Gauss method. During this calculation,
if we encounter a pivot smaller than a given value, then we reject
the corresponding function. As a result, most of the times, we cannot
use simultaneously bound and virtual states, because those two sets
span spaces which are almost identical.
In this example, we have been led to keep the 30 bound states, 3 resonances and 3 anti-resonances.
The total number of functions used to expand the solutions is then 36.

Because of the absence of pairing field, the matrix equation (\ref{eqvpga})
is equivalent to the Hartree-Fock problem:
\begin{equation}
\bigl[\mathbf T+\mathbf U\bigr]\Psi=E\mathbf R\Psi
\label{eqhf}
\end{equation}

The matrix elements of the matrices {\bf T}, {\bf R} and
{\bf U} can be computed analytically. Their expressions are given in Appendix B.
After computing the matrices in (\ref{eqhf}) and diagonalizing the matrix
{\bf R}$^{-1}$({\bf T}+{\bf U})
we obtain a set of eigenvalues and eigenvectors that we
normalize to 1 inside the domain $(0,r_0)$.

\begin{figure}[!htb]
\begin{center}
$\begin{matrix}
\includegraphics*[ scale = 0.4, angle = 0, bb = 43 48 408 302 ]{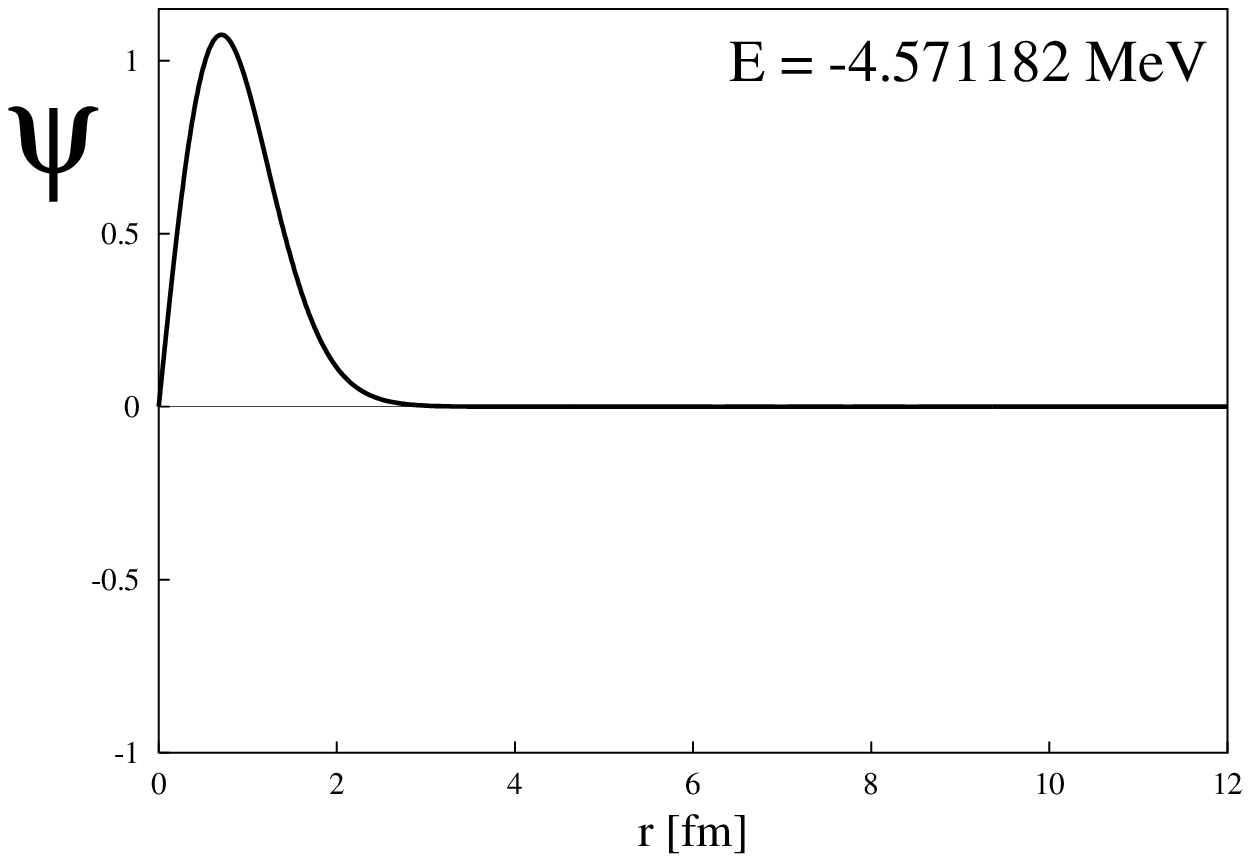} &
\includegraphics*[ scale = 0.4, angle = 0, bb = 62 48 408 302 ]{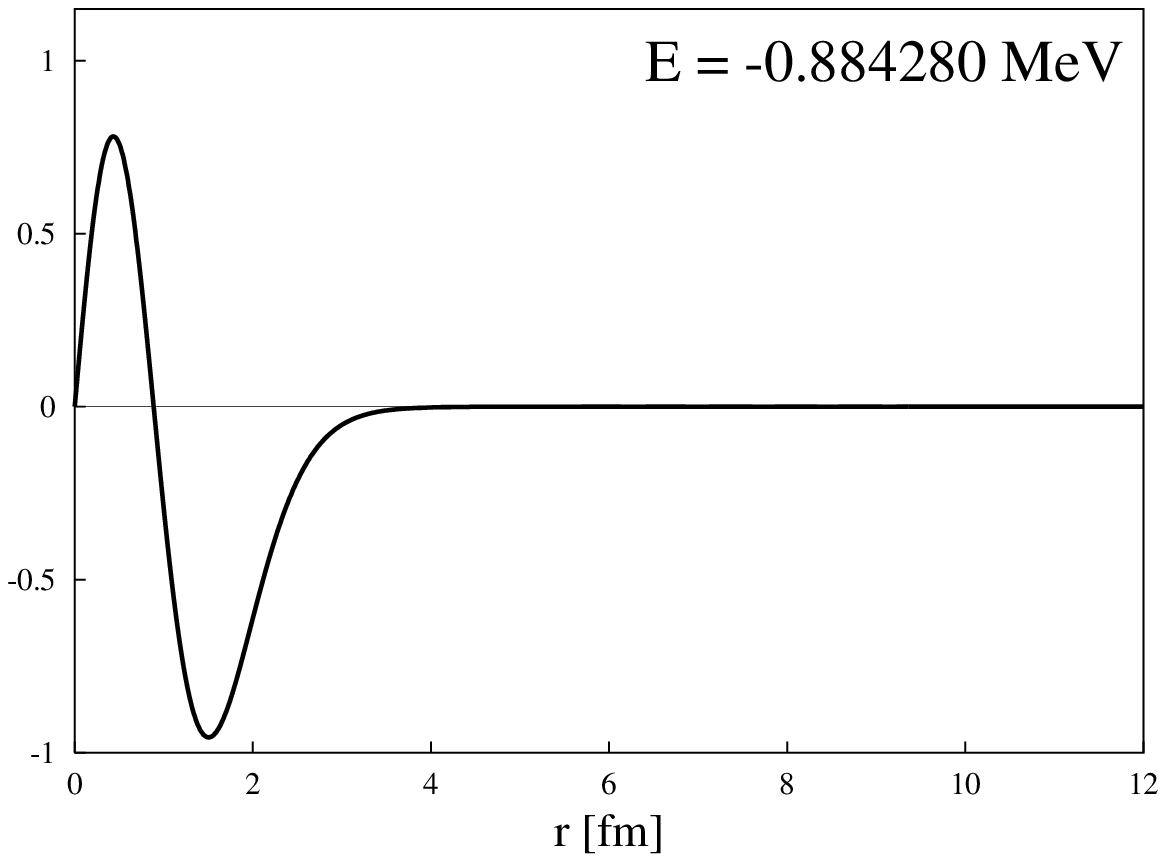} \\
\includegraphics*[ scale = 0.4, angle = 0, bb = 43 48 408 302 ]{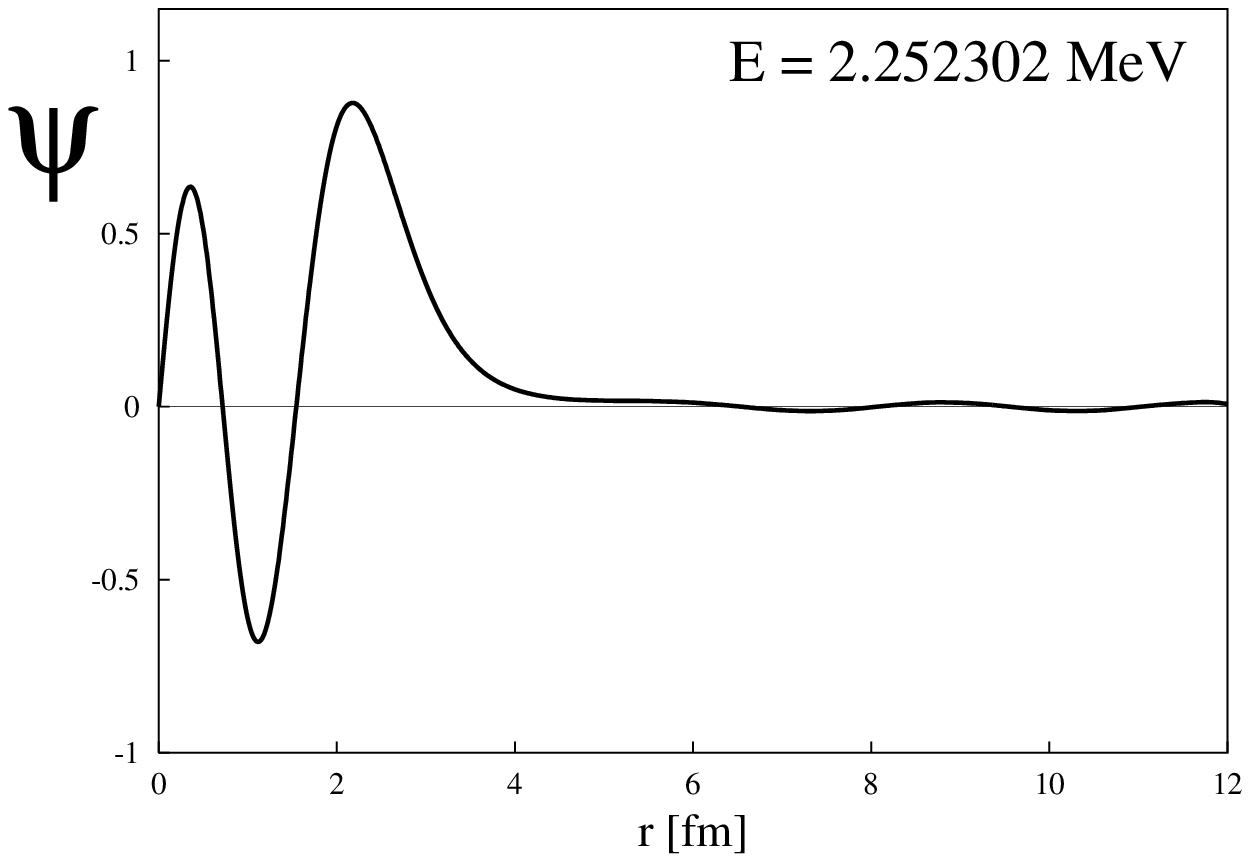} &
\includegraphics*[ scale = 0.4, angle = 0, bb = 62 48 408 302 ]{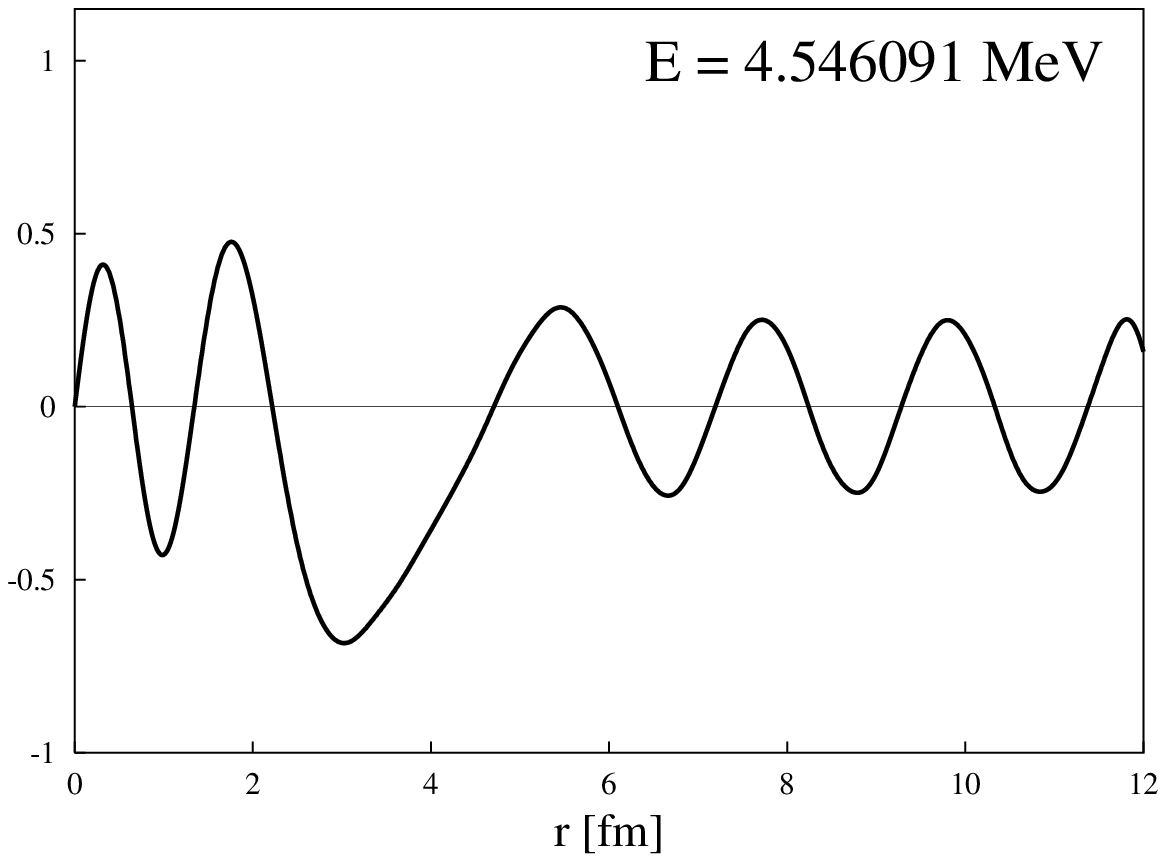}
\end{matrix}$
\caption{Wave functions obtained for $\ell=0$.
         Upper-left and Upper-right: the two bound states
         at -4.571182 MeV and -0.884280 MeV. Lower-left: the
         first resonance at 2.252302 MeV. Lower-right:
         the scattering state at 4.546091 MeV, this state
         is not much localized but can be identified as
         a resonance.
         \label{figa}}
\end{center}
\end{figure}

We have reported in table~\ref{tabb} the eigenvalues obtained below 10 MeV.
We can see that the values from table \ref{taba} are
nicely reproduced. The bound states are given with an accuracy of five to six
significant digits for the $\ell=0$ and $\ell=2$ states. The narrow resonances
are also very well reproduced. The discrepancies between the values in
table \ref{taba} and \ref{tabb} for the resonances energies are mainly
due to their widths. The wide resonances found in table \ref{taba} by computing
the S matrix cannot be seen directly in table \ref{tabb} because they spread
over a number of scattering states. The states with $\ell=1$ are not so well reproduced,
but we have checked that the agreement is better if we use
more functions to expand the solutions (in this example, all the results are
obtained by using 36 functions to expand the solutions).
A plausible explanation concerning this deficiency is the following:
the set of functions used to
expand the solutions behaves like $r$ in the limit $r\rightarrow0$, so the
convergence is supposed to be slower for partial waves with $\ell>0$, the
most important consequences are seen on waves with $\ell=1$ because
these functions have a significant amplitude near the origin where they are
not supposed to be correctly reproduced. This effect is smoothed for higher
partial waves because the centrifugal term prevent them to ``explore'' the
vincinity of the origin.
It is really remarkable to see
that the solutions with orbital angular momentum from 0 to 4 can be reproduced
by expansions on functions with $\ell=0$.

\smallskip
\begin{figure}[htbp]
\begin{center}
$\begin{matrix}
\includegraphics*[ scale = 0.365, angle = 0, bb = 40 48 408 302 ]{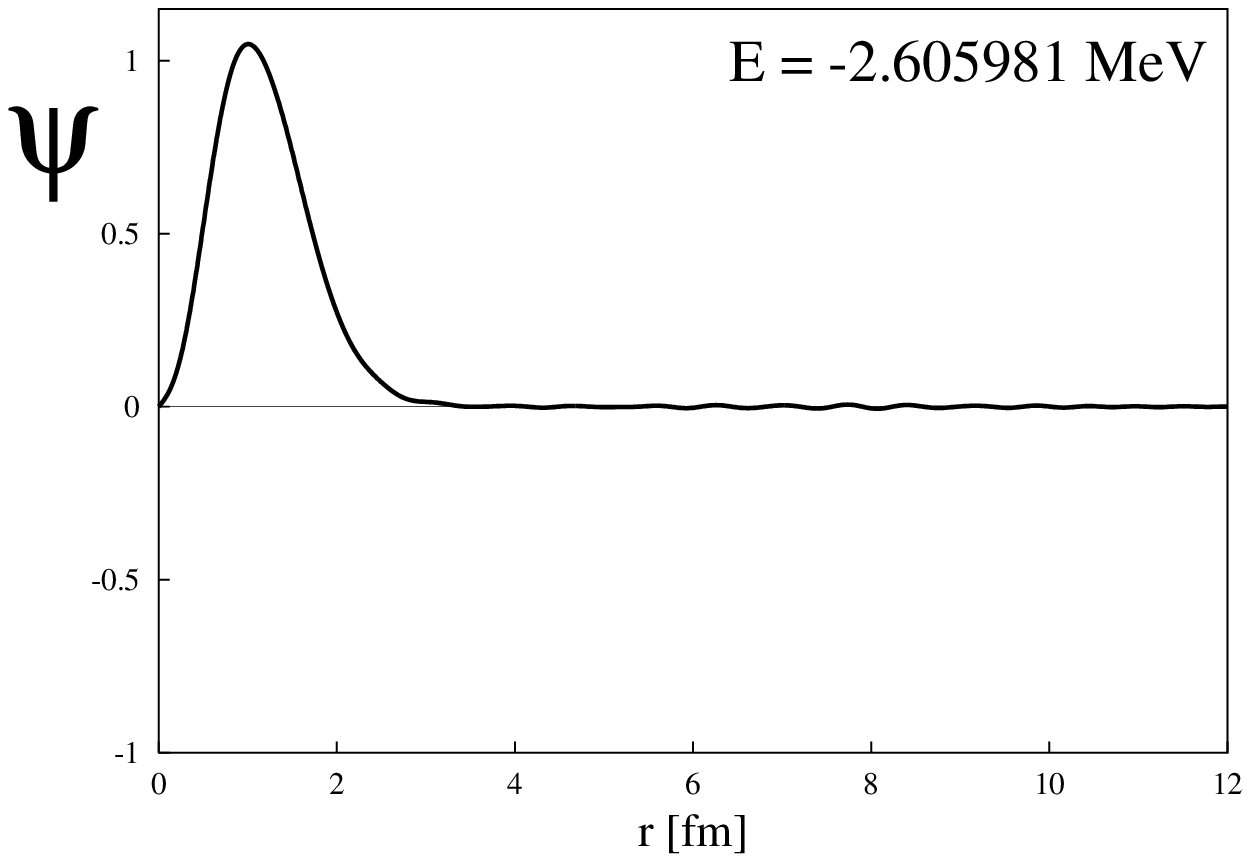} &
\includegraphics*[ scale = 0.365, angle = 0, bb = 65 48 408 302 ]{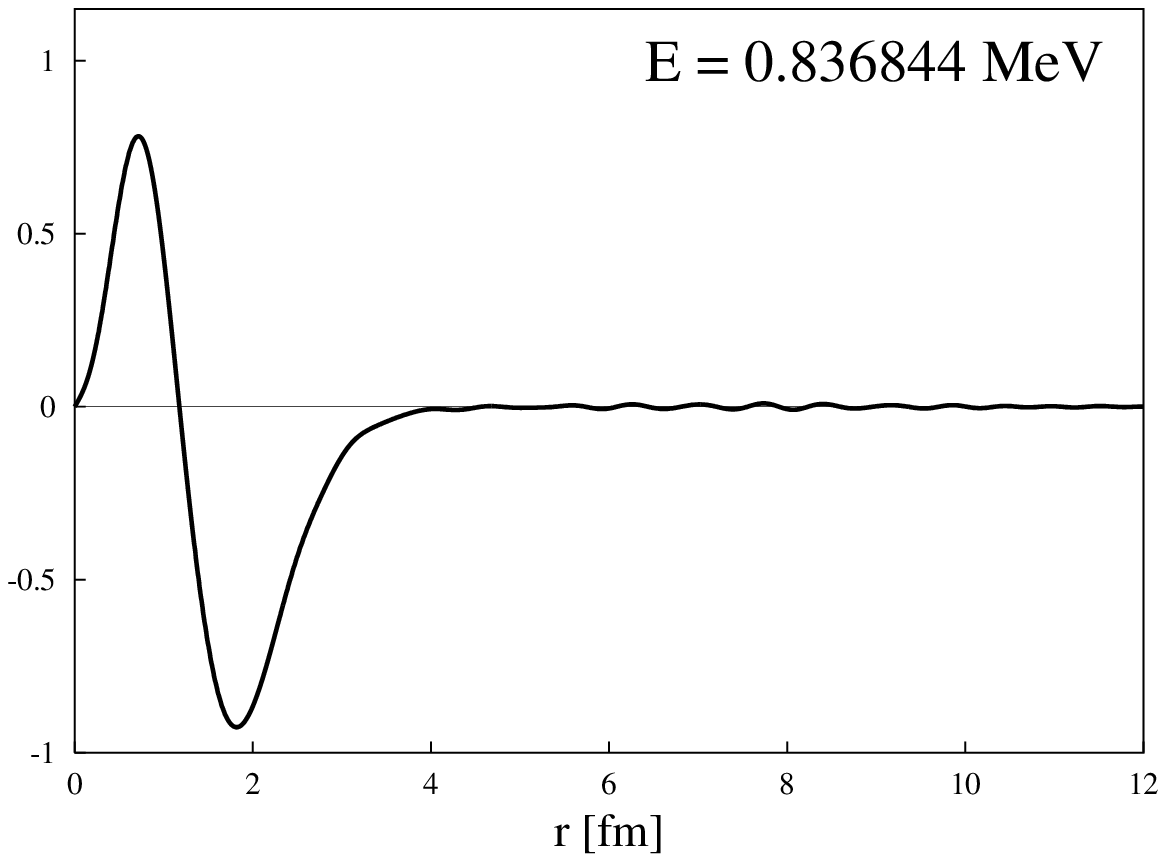} &
\includegraphics*[ scale = 0.365, angle = 0, bb = 65 48 408 302 ]{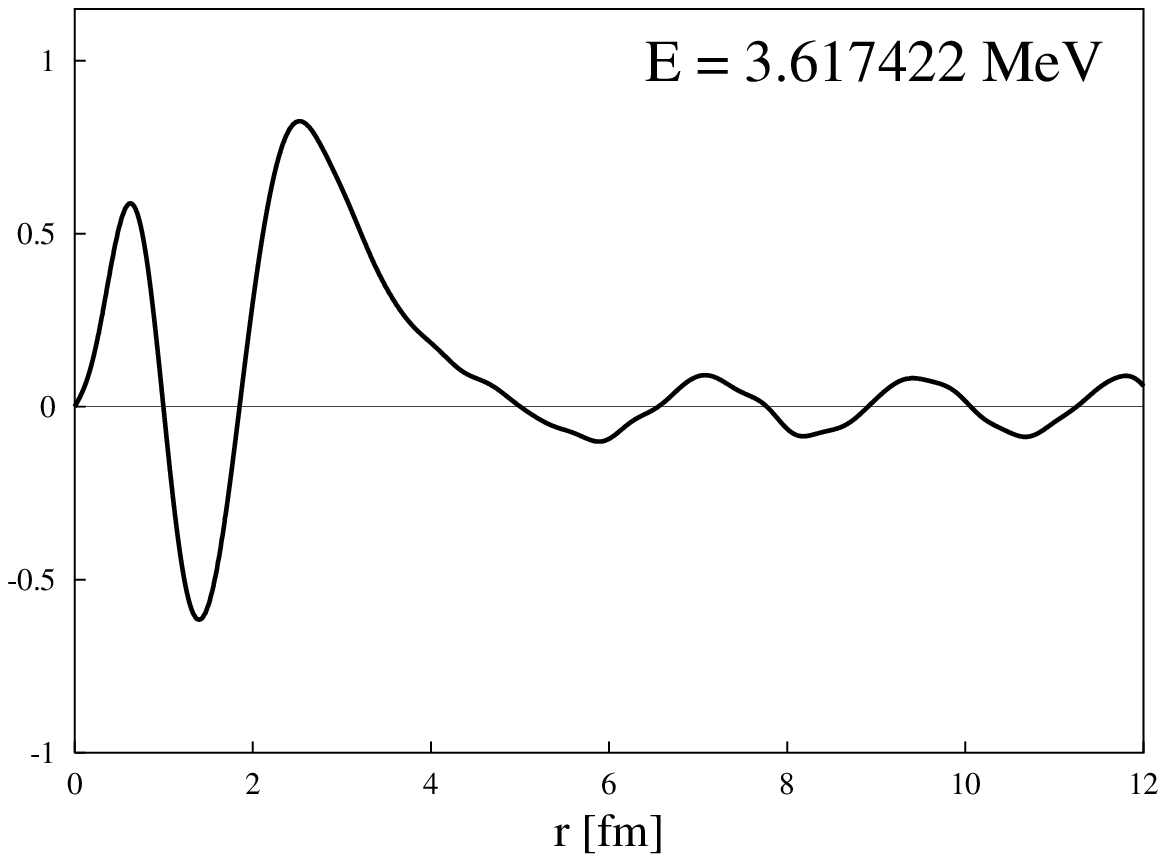}
\end{matrix}$
\caption{Wave functions obtained for $\ell=1$.
         Right: wave function of the bound state at
         -2.605981 MeV. Middle: the first resonance at 0.836844 MeV.
         Right: second resonance at 3.617422 MeV. Small unphysical
         oscillations can be seen on the wave functions, this phenomene
         is related with the behavior of the $\ell=1$ functions
         near the origin, see the text for more detailed explanations.
         \label{figaa}}
\end{center}
\end{figure}

\smallskip
\begin{figure}[htb]
\begin{center}
$\begin{matrix}
\includegraphics*[ scale = 0.365, angle = 0, bb = 40 48 400 308 ]{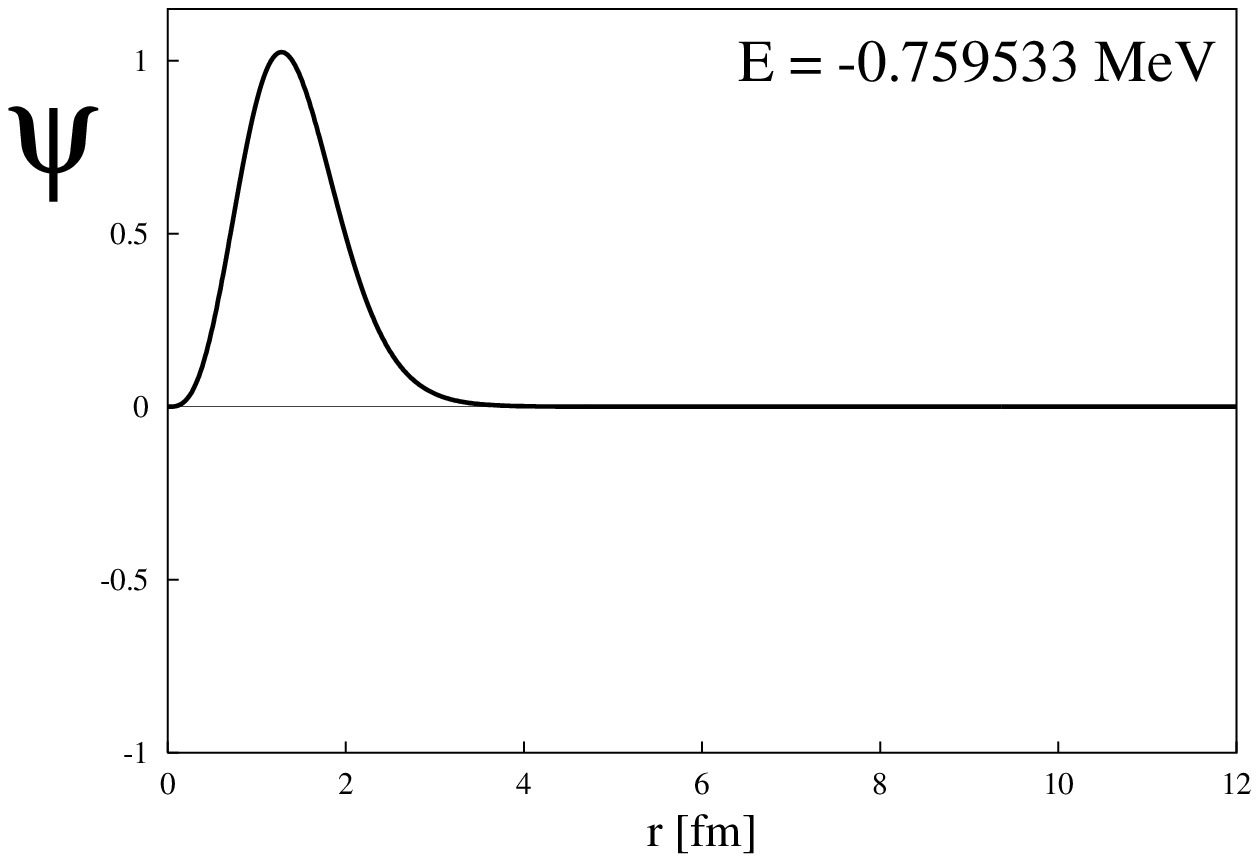} &
\includegraphics*[ scale = 0.365, angle = 0, bb = 65 48 400 308 ]{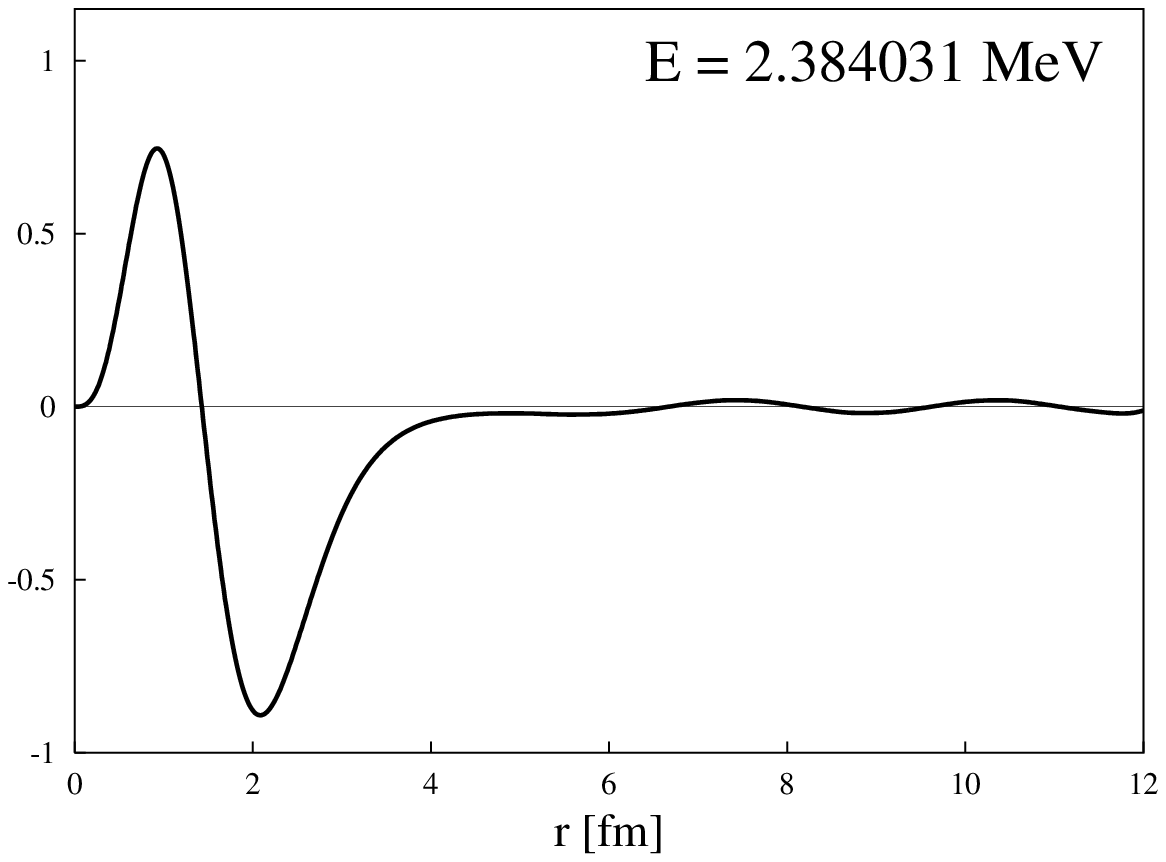} &
\includegraphics*[ scale = 0.365, angle = 0, bb = 65 48 400 308 ]{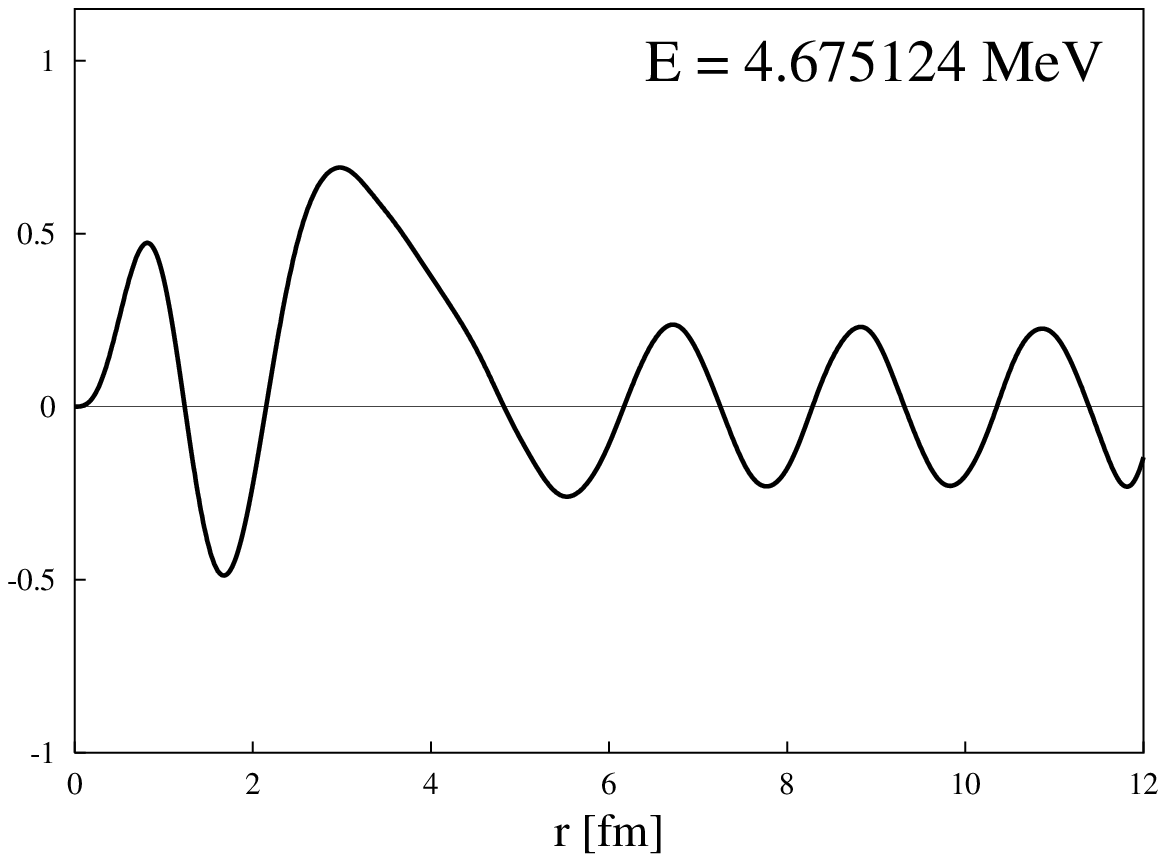}
\end{matrix}$
\caption{Wave functions obtained for $\ell=2$.
         Left: Bound state at~-0.759533 MeV.
         Middle: Resonant scattering state at 2.384031 MeV.
         Right: Resonant scattering state at 4.675124 MeV.
         The second resonance is at higher energy so the wave
         function is less localized in the interior of the
         potential than the first one.
         \label{figb}}
\end{center}
\end{figure}

We can also observe how the eigenfunctions behave. Some of them with
$\ell=0,\,1$ and 2 are plotted on figures \ref{figa},
\ref{figaa} and \ref{figb}.
Figure \ref{figa} shows that the method is able to reproduce
the deeply bound states, weakly bound states and scattering states. In the
particular case of the scattering state at 2.252302 MeV, the wave
function is strongly localized in the region of the potential and
corresponds to a resonant state.

On figure \ref{figaa} we see that the wave functions for $\ell=1$
are less correctly reproduced. Some small unphysical oscillations
appear on the tail of the functions. As it was discussed previously,
this problem can be cured by enlarging the set of functions used to
expand the solutions.

We can see on figure \ref{figb} that despite the fact that the expansion of the
solutions involves only functions with $\ell=0$, the behavior of the $\ell=0$
states at the origin ($\Psi\sim r^{\ell+1}$) is very well reproduced.
One observes that the 2.384031 MeV resonance wave function is strongly
localized in the region where the potential is attractive, while the
$4.675124$ MeV one is much less localized.

\subsection{Hartree-Fock-Bogoliubov case}

We choose ${\mathcal U}$ to be a Saxon-Woods potential:
\begin{equation}
{\mathcal U}(r)=-\frac{{\mathcal U}_0}{1+e^{\frac{r-R_0}{a}}}
\end{equation}
As the pairing field is known to be mainly localized around the nuclear
surface, we adopt for the particle-particle
channel potential ${\mathcal V}$ the derivative of a Saxon-Woods potential
with the same radius and diffuseness as in
the particle-hole channel, but with a different intensity:
\begin{equation}
{\mathcal V}(r)=\frac{d}{dr}\left[\frac{{\mathcal V}_0}{1+e^\frac{r-R_0}{a}}
\right]
\label{sawo}
\end{equation}
The system (\ref{neqa}) has been solved for partial waves
from $\ell=0$ to $4$.

As discussed in~\cite{dft}, the spectrum of (\ref{neqa}) is unbound from
above and from below. The solutions for negative
and positive energies are related by:
\begin{equation}
\left\{\begin{matrix}
  \psi_1(-E,r)=-\psi_2(E,r) \hfill \\
  \psi_2(-E,r)=\psi_1(E,r) \hfill
\end{matrix}\right.
\end{equation}
Therefore, when solving equation (\ref{neqa}), we consider only the solutions
with $E>0$.

The numerical parameters of the particle-hole potential (\ref{sawo}) have
been taken as
\begin{equation}
{\mathcal U}_0=32\mbox{\ MeV},\ \ r_0=3.7\mbox{\ fm},\ \ a=0.65\mbox{\ fm}
\end{equation}
and those of the particle-particle channel potential:
\begin{equation}
{\mathcal V}_0=4\mbox{\ MeV},\ \ r_0=3.7\mbox{\ fm},\ \ a=0.65\mbox{\ fm}
\end{equation}
The energy scale is chosen such that $\hbar^2/2m=20$ MeV in this example.

The set of functions $\{\varphi_i\}$ used to expand the solutions was
generated from a box of radius 40 fm and depth 180 MeV which contains
38 bound states and 37 virtual states. The set $\{\varphi_i\}$ is
built with the 38 bound states and the 2 first couples of resonances and
anti-resonances, so the dimension of the matrices is 42.

\begin{table}[htbp]
\begin{center}
\begin{tabular}{|l|r|r|r|r|}
\hline
\hline
$\ell$     &        0 &      0 &      1 &      2 \\
\hline
 E (MeV) \ &  -19.288 & -0.856 & -9.460 & -0.124 \\
\hline
\hline
\end{tabular}
\caption{Energies of the bound states in the Saxon-Woods potential
         (without pairing).
         \label{tabnrjws}}
\end{center}
\end{table}

In table \ref{tabnrjws} we have reported the energies and the norms of the particle-particle bound states when the pairing field and the
chemical parameter are
set to zero. The system has one loosely bound state 124 keV below
the continuum. As mentioned in the previous paragraph,
the convergence in the basis of the different partial waves is quite fast and good
except for $\ell=1$ for which it is slower. For this reason, we will
present in detail the results only for $\ell$=0 and 2. As in the previous
paragraph, we have checked that a satisfactory convergence for $\ell=1$ can
be obtained by enlarging the set of functions $\{\varphi_i\}$.
In a future development of this method, where most general case of 3 dimensionnal
computation will be investigated, this problem will not appear because
the effective centrifugal potential is not present anymore.

\begin{figure}[htb]
\begin{center}
$\begin{matrix}
\includegraphics*[ scale = 0.365, angle = 0, bb = 40 48 400 302 ]{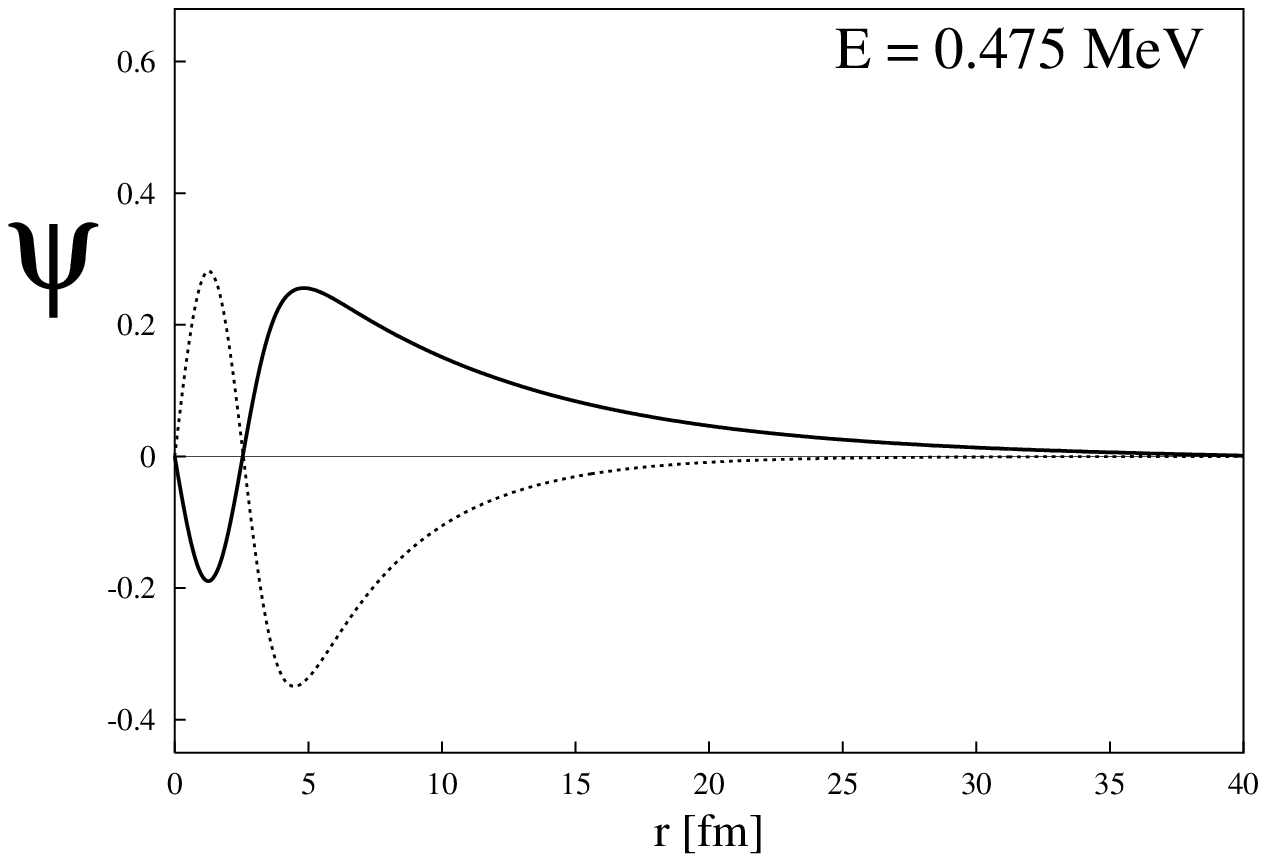} &
\includegraphics*[ scale = 0.365, angle = 0, bb = 65 48 400 302 ]{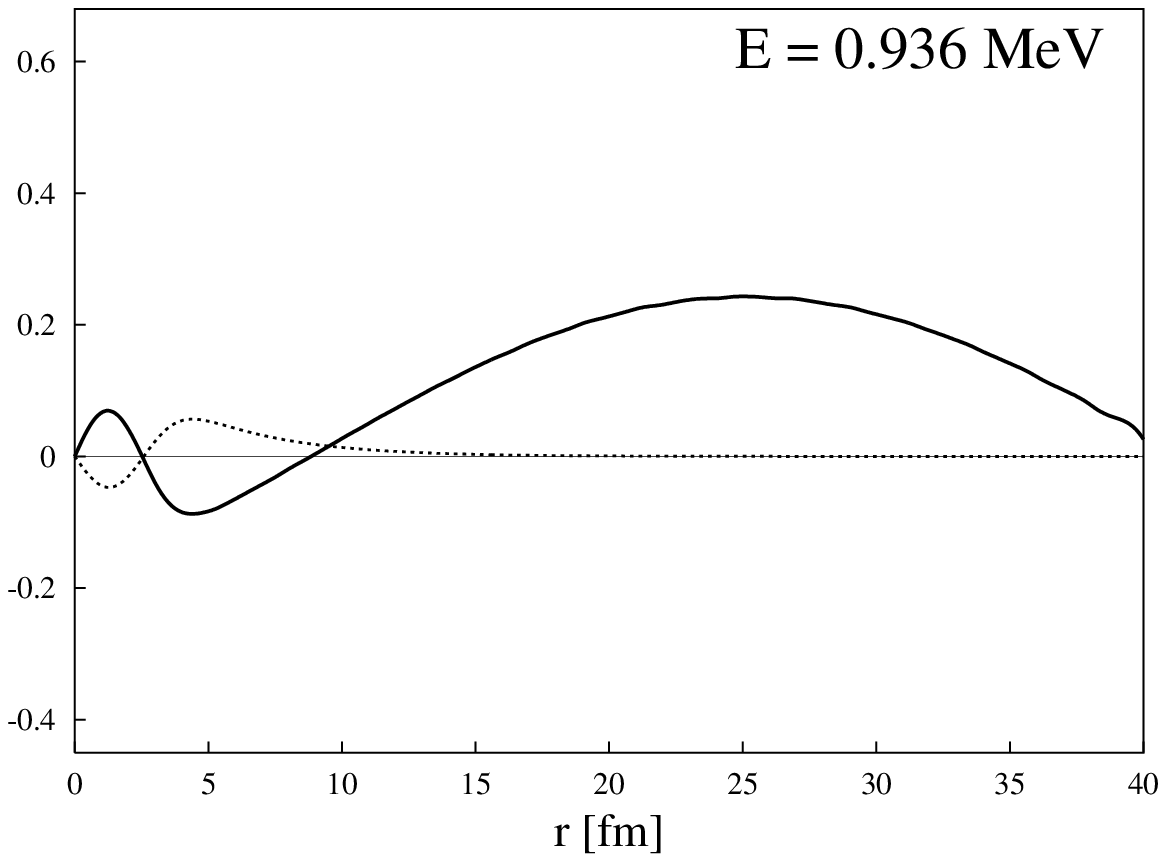} &
\includegraphics*[ scale = 0.365, angle = 0, bb = 65 48 400 302 ]{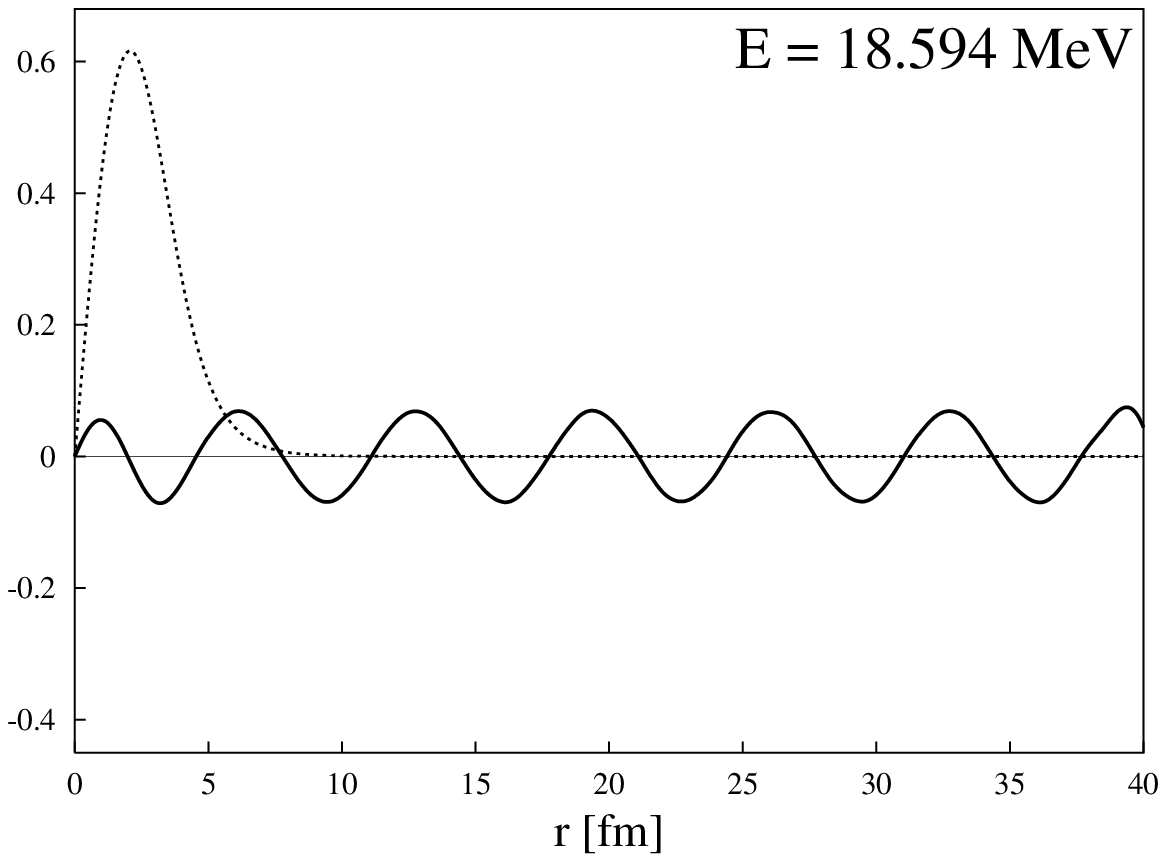}
\end{matrix}$
\caption{Wave functions of three quasi-particle states for $\ell=0$.
         The two components $\psi_1$ (solid line) and $\psi_2$ (dashed line)
         have the same asymptotic behaviour for the state on the left panel
         which corresponds to a discret level ($E<-\lambda$), while different
         asymptotic behaviours are obtained for the two others which
         correspond to states with quasi-particle energies $E>-\lambda$.
         \label{figqp}}
\end{center}
\end{figure}

In table \ref{tabnrjwsqp} we have reported the energies and the norms $N_2$ of the lower components of the quasi-particle states for which
$N_2$ is larger than $0.001$,
for $\ell=0$ and $\ell=2$. For higher orbital momenta, the lower component of all solutions is almost zero and does not contribute to
the particle density. With the value chosen for the chemical potential ($\lambda = -0.750$ MeV), the number of particles in the
system is on the average
3.163 for $\ell=0$ and 0.417 for $\ell=2$. We have represented in figure \ref{figqp} three quasi-particle wave functions:
two states close to the Fermi sea and a deep hole state, with quasi-particle energies 0.475 MeV, 0.936 MeV and 18.594 MeV, respectively.
The energy of the first quasi-particle
state is smaller than $|\lambda|$ and therefore this state belongs to the discrete spectrum (this is actually the only discrete state in the full
spectrum). The asymptotic behaviour of the quasi-particle wave function
is~\cite{dft}:
\begin{equation}
\left\{\begin{matrix}
  \ds \psi_1(r)\propto \exp(-\kappa_1r) \hfill \\
\noalign{\vskip 1mm}
  \ds \psi_2(r)\propto \exp(-\kappa_2 r) \hfill
\end{matrix}\right.
\ \ \ \ \ \mbox{for}\ r\rightarrow\infty.
\end{equation}
The two other quasi-particle states shown in the figure are continuum states with an asymptotic behaviour given by:
\begin{equation}
\left\{\begin{matrix}
  \ds \psi_1(r)\propto \sin(k_1 r)  \hfill \\
\noalign{\vskip 1mm}
  \ds \psi_2(r)\propto \exp(-\kappa_2 r) \hfill 
\end{matrix}\right.
\ \ \ \ \ \mbox{for}\ r\rightarrow\infty,
\end{equation}
with obvious definitions for $k_1$, $\kappa_1$ and $\kappa_2$.

\begin{figure}[htb]
\begin{center}
$\begin{matrix}
\includegraphics*[ scale = 0.5, angle = 0, bb = 35 48 410 302 ]{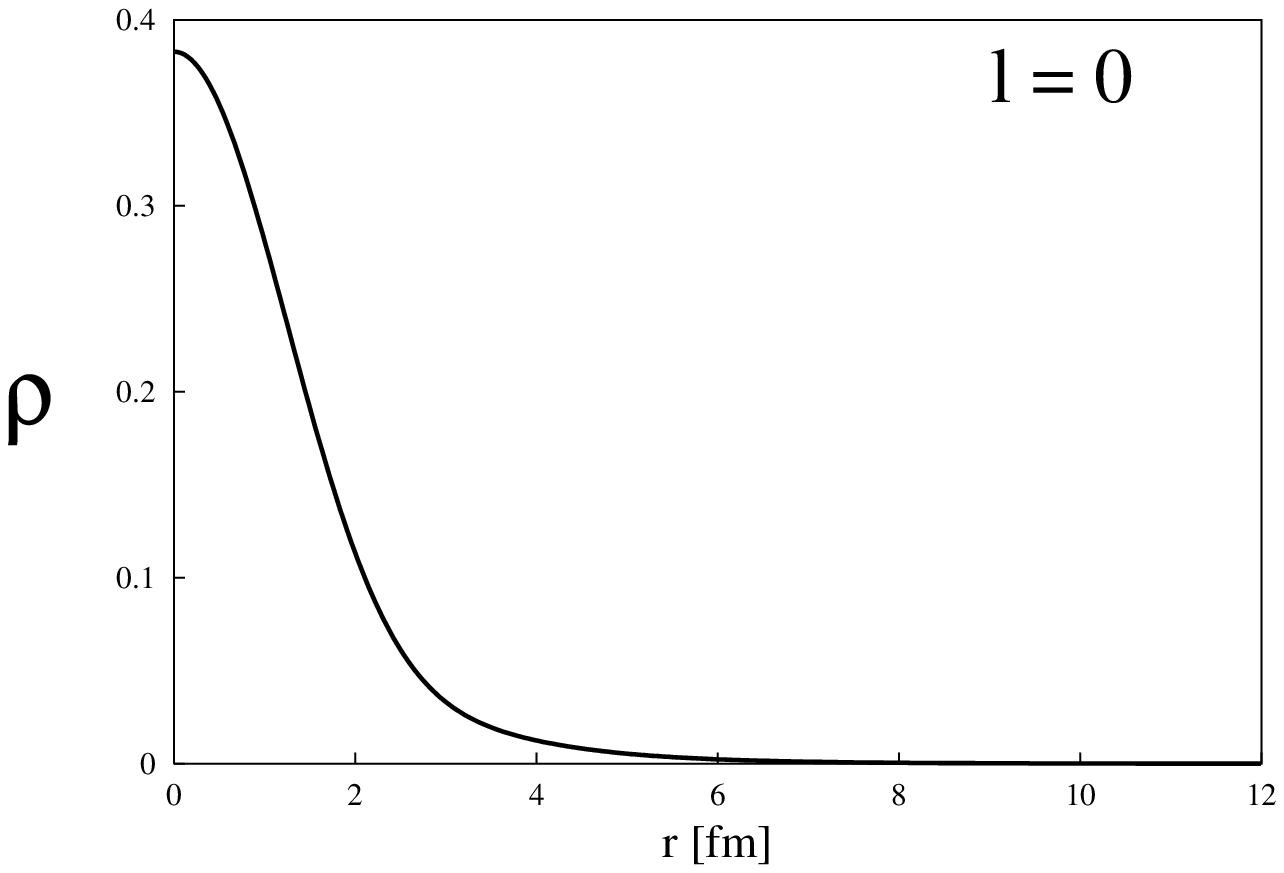} &
\includegraphics*[ scale = 0.5, angle = 0, bb = 50 48 410 302 ]{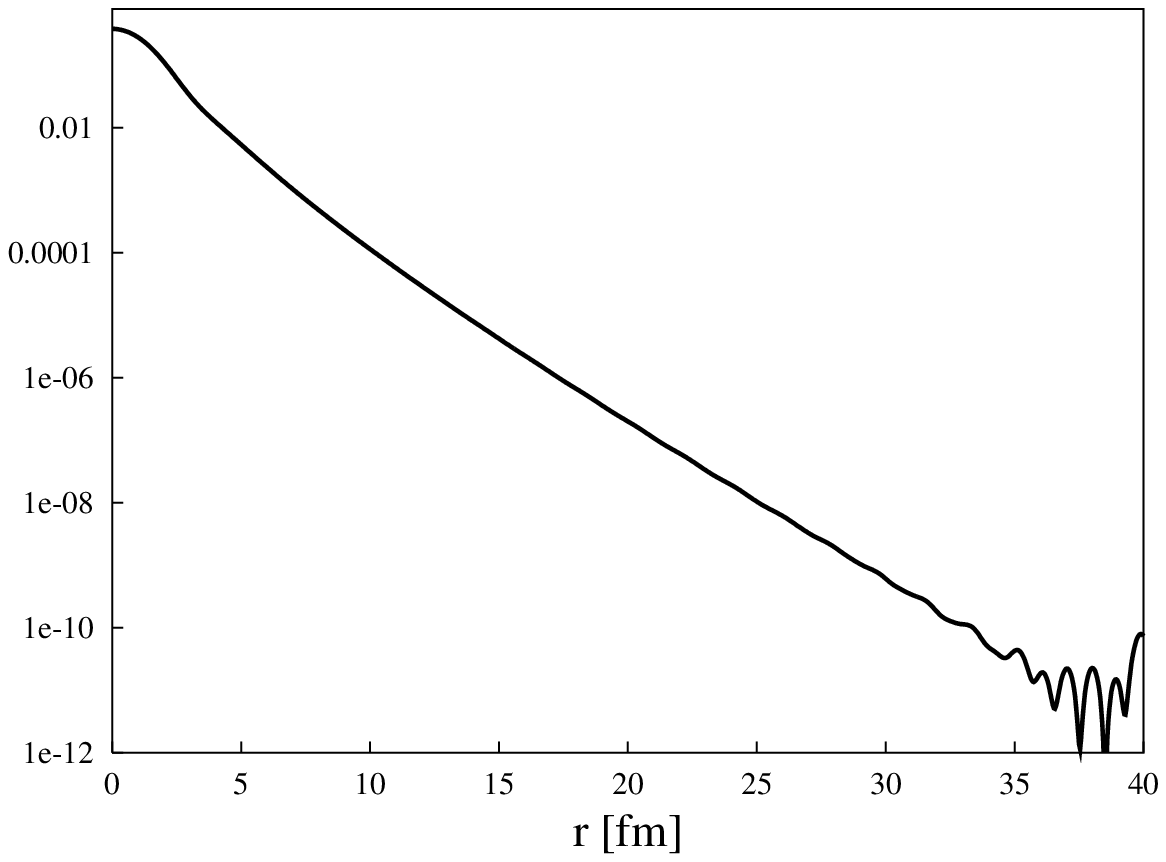} \\
\includegraphics*[ scale = 0.5, angle = 0, bb = 35 48 410 302 ]{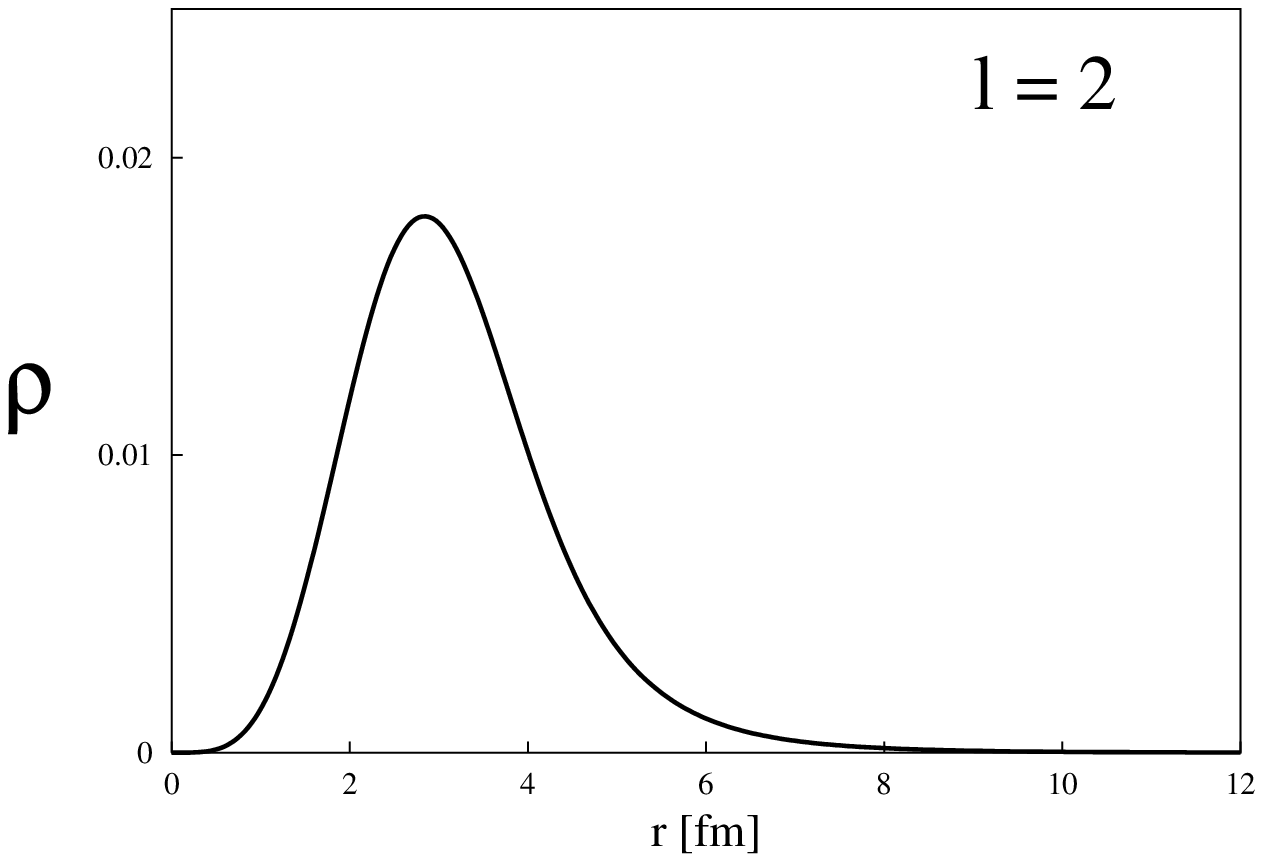} &
\includegraphics*[ scale = 0.5, angle = 0, bb = 50 48 410 302 ]{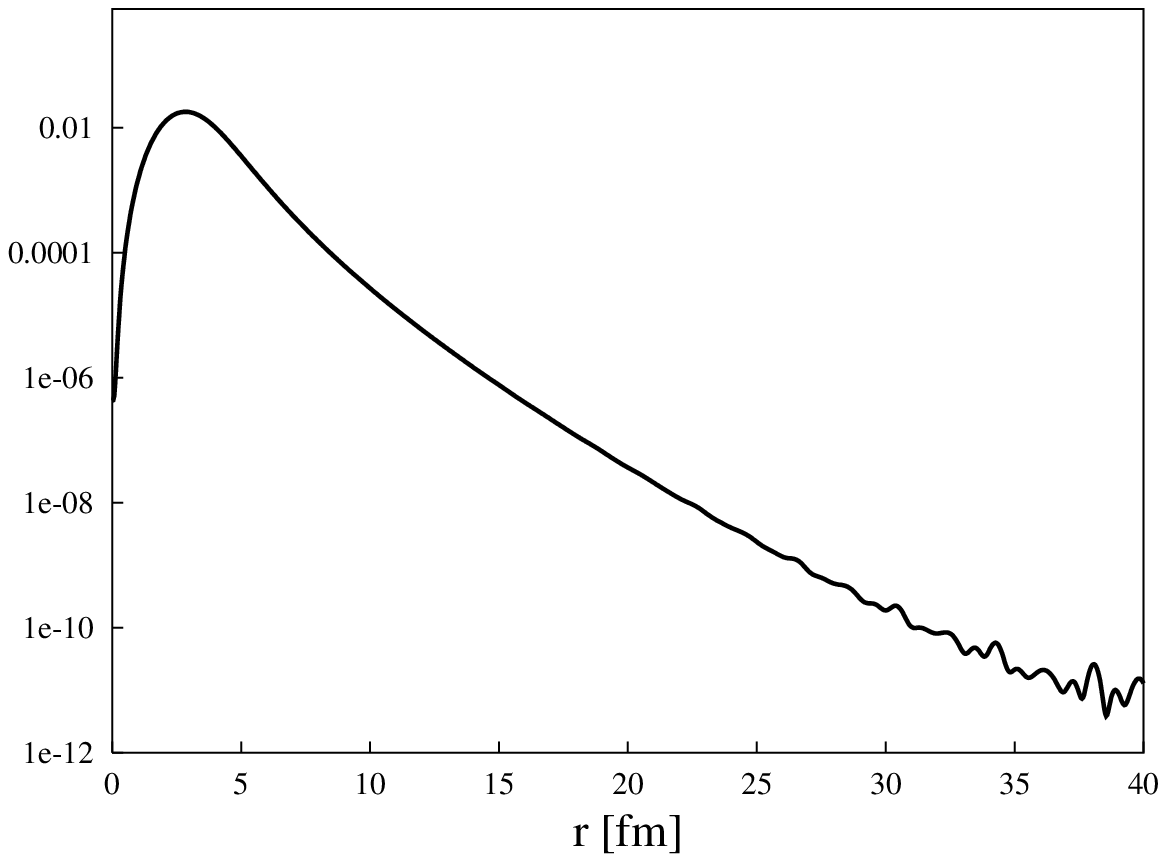}
\end{matrix}$
\caption{Particle densities in fm$^{-3}$ for $\ell=0$ (top) and $\ell=2$
         (bottom) as functions of~$r$.
         Figures on the right hand side are in logarithmic scale.
         \label{figc}}
\end{center}
\end{figure}

We can see in figure \ref{figqp} that the method we employ here nicely reproduces those two completely different asymptotic behaviours.
\begin{table}[htbp]
\begin{center}
\begin{tabular}{|r|c|r|c||r|c|}
\hline
\hline
\multicolumn{4}{|c||}{$\ell=0$}&\multicolumn{2}{c|}{$\ell=2$} \\
\hline
 E (MeV) & $N_2$ & E (MeV) & $N_2$ & E (MeV) & $N_2$ \\
\hline
  0.475 \ & \ 0.549 \ &  4.662 \ & \ 0.001 \ & 1.072 \ & \ 0.152 \ \\
\noalign{\vskip -2pt}
  0.936 \ &   0.013   & 15.338 \ &   0.001   & 1.157 \ &   0.043   \\
\noalign{\vskip -2pt}
  1.443 \ &   0.010   & 18.212 \ &   0.090   & 1.742 \ &   0.004   \\
\noalign{\vskip -2pt}
  2.239 \ &   0.005   & 18.594 \ &   0.905   & 2.587 \ &   0.002   \\
\noalign{\vskip -2pt}
  3.315 \ &   0.002   & 21.402 \ &   0.001   & 3.689 \ &   0.001   \\
\hline
\hline
\end{tabular}
\caption{Energies E and norms $N_2$ of the lower components of the
         quasi-particle
         states with orbital momenta $\ell$=0 and 2. Only the states with
         $N_2>0.001$ have been reported.
         \label{tabnrjwsqp}}
\end{center}
\end{table}
Using the relation given in equation (\ref{norden}) the densities of particles for $\ell$=0 and 2 can be built.
These quantities are represented in figure \ref{figc} using linear and logarithmic
scales.

We notice on the logarithmic plots that the behaviour of the density becomes oscillatory when it falls below $10^{-9}$ fm$^{-3}$. This is
due to the truncation of the basis expansion and also to some numerical inaccuracies.
However the particle density is so small in this region that this pathology should not play any significant role in the calculation of
any observable.

\begin{figure}[htb]
\begin{center}
$\begin{matrix}
\includegraphics*[ scale = 0.365, angle = 0, bb = 35 48 400 302 ]{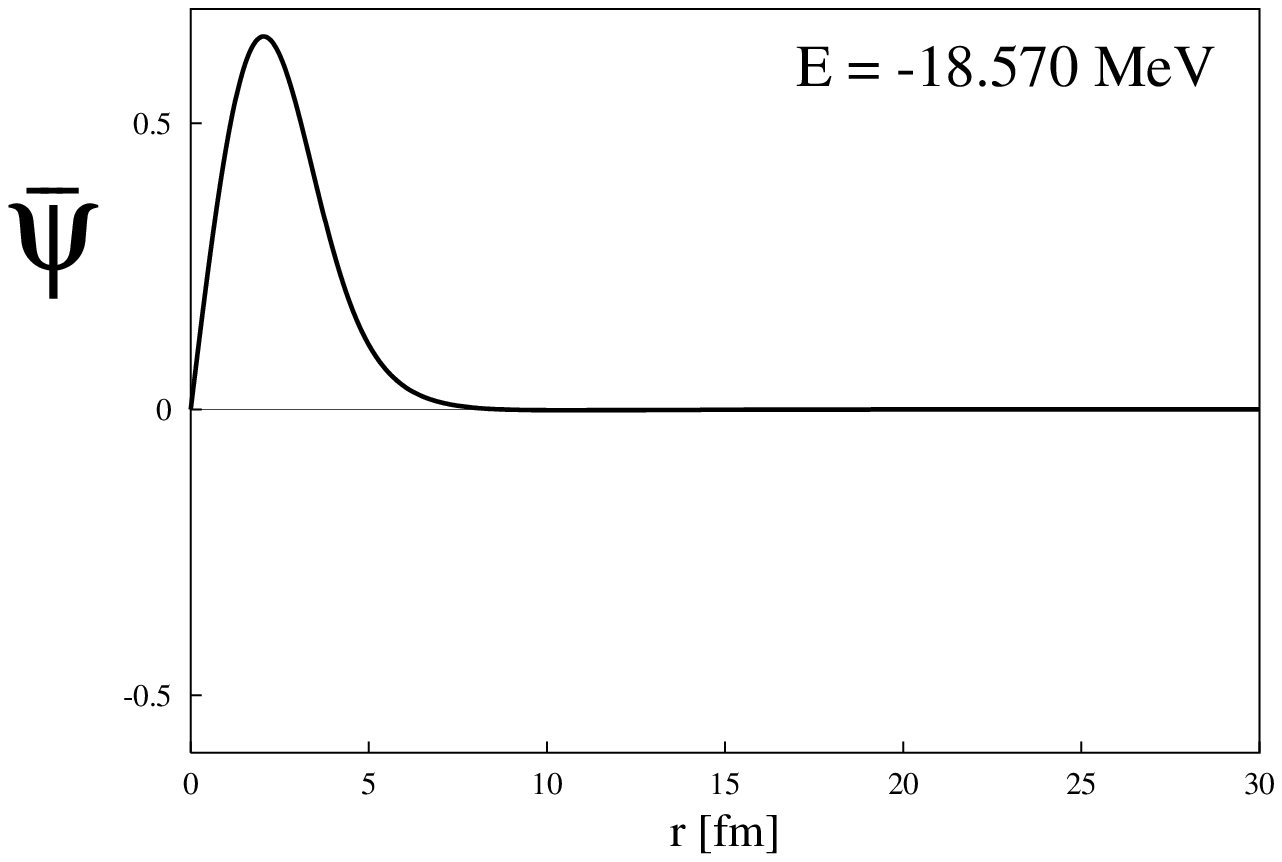} &
\includegraphics*[ scale = 0.365, angle = 0, bb = 65 48 400 302 ]{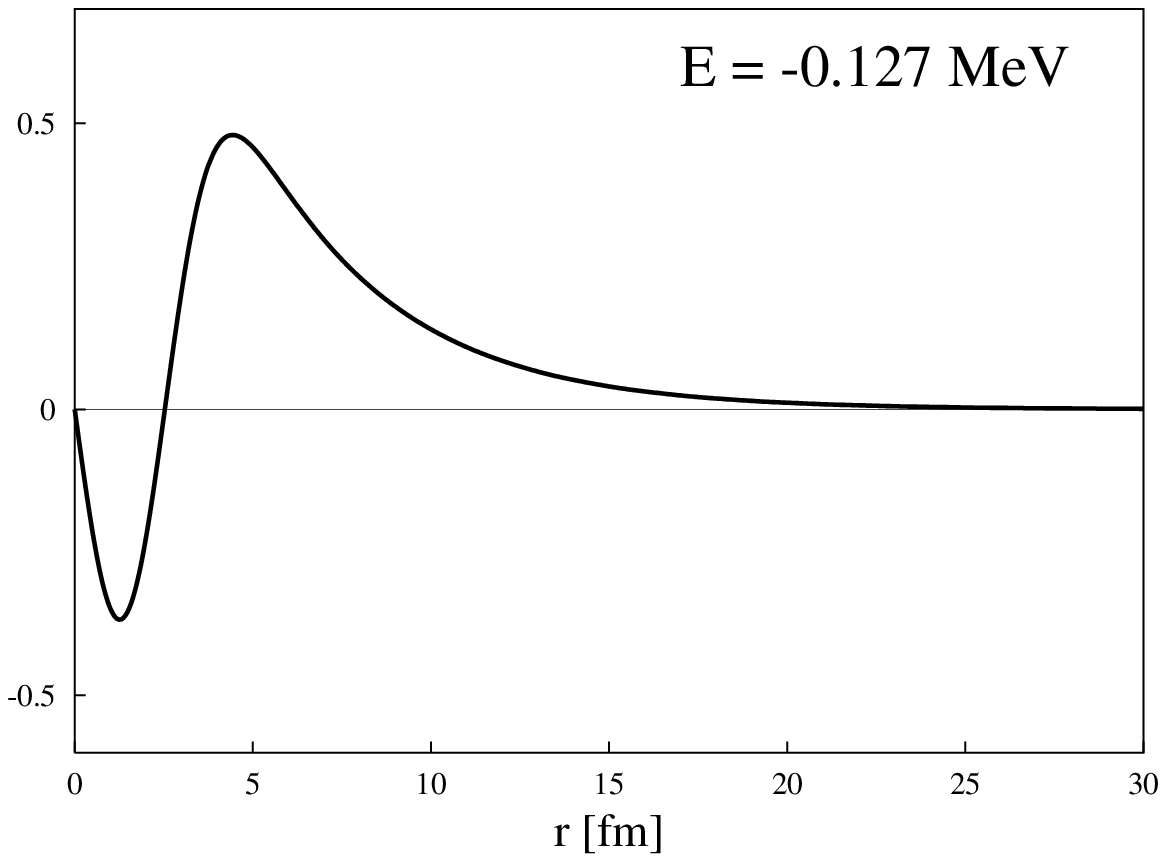} &
\includegraphics*[ scale = 0.365, angle = 0, bb = 65 48 400 302 ]{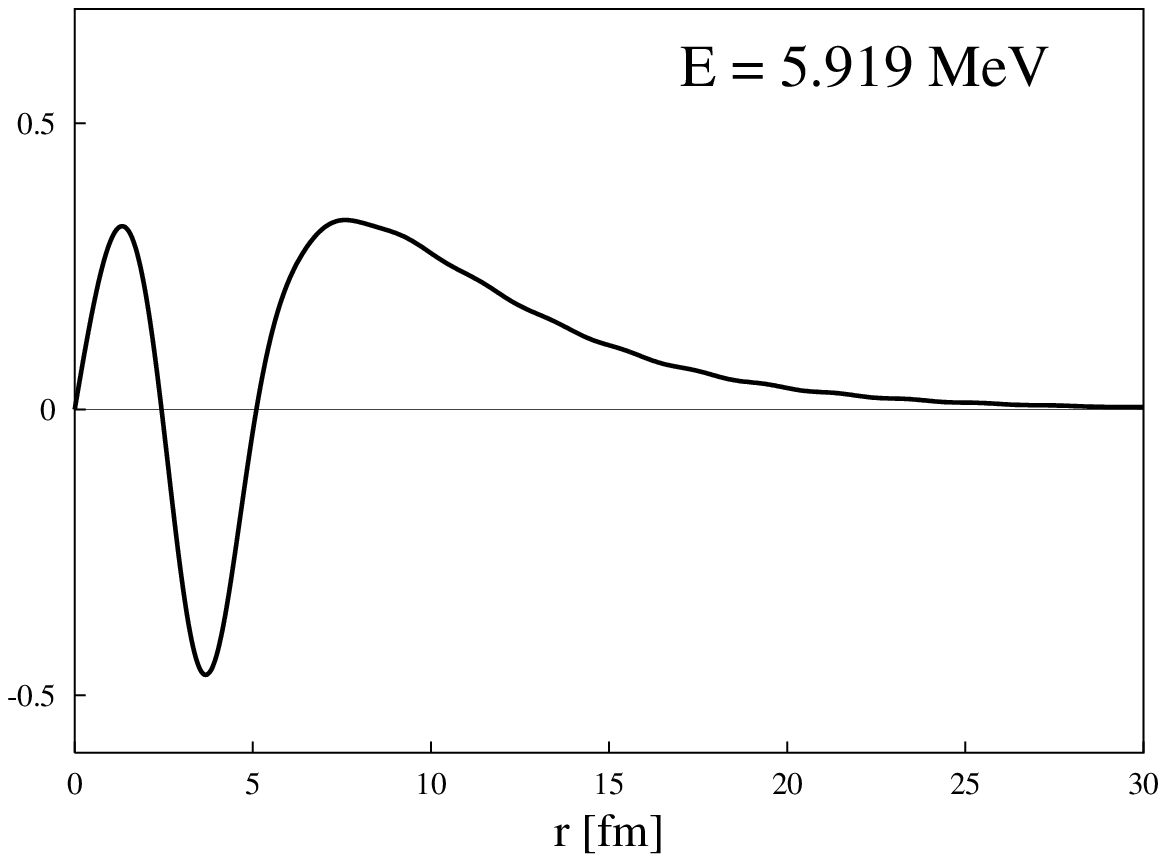}
\end{matrix}$
\caption{Wave functions and energies
of the three first canonical states for $\ell=0$.
         \label{figd}}
\end{center}
\end{figure}

Eventually we diagonalize the density operator in order to construct the
canonical
states. The single-particle energies and occupations of $\ell=0$ canonical
states are displayed in table \ref{tabcan}.
The wave-functions of the first three ones are shown in figure \ref{figd}.
We notice that the expected decrease to zero at large $r$ of these wave
functions is remarkably well described within the present approach.

\begin{table}[htbp]
\begin{center}
\begin{tabular}{|l|r|r|r|r|}
\hline
\hline
$\epsilon$ (MeV) & -18.570  & -0.127  & 5.919  & 24.570  \\
\hline
$v^2$     & \ \  0.9997 & \ \ 0.5802 & \ \ 0.0012 & \ \ 0.0002 \\
\hline
\hline
\end{tabular}
\caption{Energies $\epsilon$ and occupations $v^2$ of the canonical states
         obtained for $\ell=0$. Only the states with an occupation greater
         than 0.0001 have been reported.
         \label{tabcan}}
\end{center}
\end{table}

\section{Conclusion}
\label{secd}

We have presented in this work an expansion method which can be used to
express both discrete and continuum solutions of the most general HF or HFB equations, when any kind of nucleon-nucleon interaction is used.
Because of the simplicity of the functions on which we expand the solutions, many steps of the resolution can be done analytically. This
is the great advantage of the method. For example, the matrices needed for the resolution of the equations (\ref{eqvpga}) can be evaluated
analytically if the mean-field potentials are of Saxon-Woods, and derivative of Saxon-Woods or Gaussian forms.

The fact that this method
avoids the problems associated with the discretization of integro-differential equations has an important consequence: it is fast and accurate.
The only part of the problem that has to be done numerically is the inversion of
the matrix ${\mathbf R}$ and the diagonalisation of (\ref{eqvpga}).
In general, the dimension involved are rather small, at least in the case of spherical symmetry. So these steps can be
done rapidly and with a good accuracy.
In addition, as the quasi-particle states are expressed as finite expansions on a basis of wave-functions, the method ensures that the
variational HF and HFB procedures lead to an upper bound of the total energy of the system, which can be improved at will by increasing
the size of the basis.

Because of the form of the kinetic energy (\ref{eqemt}), the matrix problem that we solve
is formally equivalent to solving the corresponding integro-differential equations in
a box of radius $r_0$ with the non local boundary condition
$\Psi'(r_0)=iK\Psi(r_0)$
for the solutions, where $K$ is the
operator defined by:
$$\int\!\varphi^*_m(r)K\varphi_n(r)\,dr=k_n\delta_{mn},$$
and $k_n$ are the solutions of equation (\ref{treq}).
In this sense, the present method is completely equivalent to working directly in
the coordinate representation as done e.g. in~\cite{dft} and,
provided the size of the box is the same, comparable results would be obtained.
The advantage of the technique we propose is that it can be
applied to HFB calculations employing a finite-range nuclear interaction such as the Gogny force. Actually, once the matrix elements of the
two-body force are computed, the method allows one to solve the HF and HFB equations as easily as with a basis of harmonic oscillator states,
which has been done extensively with the Gogny force~\cite{dg,gg}.
In addition, extensions of the method to non-spherical nuclei could
be done in the same fashion as with harmonic oscillator bases.

Let us note that in the limit where the depth of the rectangular well goes to infinity ($V_0\rightarrow+\infty$ in equation (\ref{pui}))
usual discrete Fourier expansions are recovered. In that case the formalism is greatly simplified because the transcendental equation
(\ref{treq}) becomes trivial and the overlap matrix {\bf R} reduces to unity.

The more general case investigated here has however the advantage of defining a reference set of functions which are  adapted to the
expansion of the (quasi-)particle state solutions of the HFB equations in nuclear systems. As shown in the present work, good approximation
of the solutions can therefore be obtained, using comparatively small expansions.

The implementation of the present technique in the fully self-consistent HFB procedure will be the next step of this study. An important issue
will be to check if nuclear properties can be accurately reproduced in realistic situations, in particular in nuclei close to drip lines, by
using bases small enough to ensure extensive calculations even in heavy nuclei. If such a requirement is met, the method of solving the HFB
equations we propose should open a broad range of new nuclear studies for the future.


\appendix

\section{The decomposition of identity}

In this paragraph, we consider $\hbar^2/2m=1$ to simplify the expressions.
We consider the Green function $G^{KP}(r,r';k)$ (KP for Kapur-Peierls),
corresponding
to mixed boundary conditions on the domain $]0,r_0[$, solution of the (non
self adjoint) problem:
\begin{equation}
{ \left\{ \begin{array}{ll}
     \ds\left[-\frac{d^2}{dr^2}+(V(r)-k^2)\right]G^{KP}(r,r';k)=\delta(r-r'),
           & \mbox{if}\ 0<r,r'<r_0  ,\\[2mm]
                     G^{KP}(0,r';k)=0 , & \\[2mm]
     \ds \frac{d}{dr}G^{KP}(r,r';k)|_{r=r_0}=
          i k G^{KP}(r,r';k)|_{r=r_0}& \\
                      \end{array}\right.}
\label{gkp}
\end{equation}
If we put $p=\sqrt{k^2+V_0}$, the solution reads:
\begin{equation}
{\displaystyle \left\{ \begin{array}{c}
                        G^{(1)}(r,r';k)=A_1e^{i pr}+B_1e^{-\imath pr}  ,
                  \ \mbox{for}\ 0\leq r\leq r'  ,\\[2mm]
                        G^{(2)}(r,r';k)=A_2e^{i pr}+B_2e^{-\imath pr}  ,
                  \ \mbox{for}\ r'\leq r\leq r_0 ,
                      \end{array}\right.}
\label{gkpir}
\end{equation}
The coefficients are given by the boundary conditions:
\begin{equation}
{\displaystyle \left\{ \begin{array}{l}
                        G^{(1)}(0,r';k)=0 ,\\[2mm]
                        G^{(2)}(r',r';k)=G^{(1)}(r',r';k) ,\\[2mm]
                        (G^{(2)})'(r'_+,r';k)=(G^{(1)})'(r'_-,r';k)-1 ,\\[2mm]
                        (G^{(2)})'(r_0,r';k)= i k G^{(2)}(r_0,r';k) ,
                      \end{array}\right.}
\label{bcoir}
\end{equation}
and we obtain:
\begin{equation}
{\displaystyle  G^{KP} (r,r';k)=\left| \begin{array}{c}
         \ds-\frac{i k\sin p(r'-r_0)+p\cos p(r'-r_0)}
{i k \sin pr_0-p \cos pr_0}
\frac{\sin pr}{p},\ \ \ \mbox{if}\ 0\leq r\leq r',\\[5mm]
         \ds-\frac{i k\sin p(r-r_0)+p\cos p(r-r_0)}
{i k \sin pr_0-p \cos pr_0}
\frac{\sin pr'}{p},\ \ \ \mbox{if}\ r'\leq r\leq r_0,
                      \end{array}\right.}
\label{soluir}
\end{equation}
One checks that $k\rightarrow G^{KP}(r,r';k)$ can be meromorphically
continued in the complex
plane and that for $0<r,r'<r_0$, it decays exponentially when
$|k|\rightarrow \infty$ in the
whole complex plane ${\mathbb C}$.

\smallskip

In order to get a completeness formula, we consider a compactly supported
$C^2$ function $f$,
(which is zero near the end points $r=0$ and $r=r_0$, and we denote by $g$
the function defined by:
\begin{equation}
{\displaystyle  g(r)=\left( -\frac{d^2}{dr^2}-V_0 \right)f(r).}
\label{g}
\end{equation}
By multiplying (\ref{gkp}) by $f$, (\ref{g}) by $G^{KP}(r,r';k)$, subtracting
the two resulting
equations and integrating over $(0,r_0)$, we get:
\begin{equation}
{\displaystyle
\int_0^{r_0} kG^{KP}(r,r';k)f(r)\ dr=-\frac{1}{k}f(r')+
\frac{1}{k}\int_0^{r_0} G^{KP}(r,r';k)g(r)\ dr.}
\label{gr}
\end{equation}
By integrating on each side on a big circle $C_{\Lambda}=\{k\in
{\mathbb C}\ :\ |k|=\Lambda \}$, we obtain:
\begin{equation}
{\displaystyle
\lim_{k\rightarrow \infty}\frac{1}{2i \pi}\int_{C_{\Lambda}}
\left[\int_0^{r_0} k G^{KP}(r,r';k)f(r)\ dr\right]\ dk=-f(r').}
\label{gr1}
\end{equation}

\g By applying the Cauchy residues theorem, we get the following formula:
\begin{equation}
{\displaystyle
f(r')=-\sum_n Res \left[\int_0^{r_0} k G^{KP}(r,r';k)f(r)\ dr\right]_{k=k_n},}
\label{gr2}
\end{equation}
where the $k_n$ are the solutions of the equation :
\begin{equation}
p_n\cot p_nr_0=i k_n,
\label{trans}
\end{equation}
with $p_n=\sqrt{k_n^2+V_0}$
After computing the residue, we get the following decomposition identity:
\begin{equation}
{\displaystyle
f(r')=
\sum_n
\frac{k_n}{i + k_nr_0}
\left(\int_0^{r_0} \sin p_nr    f(r)\ dr \right) \sin p_nr'}.
\label{gr500}
\end{equation}
In this identity, the index $n$ labels all the complex solutions of
(\ref{trans}).

One may assume for convenience that the first $N$ roots are pure imaginary
$k_n=i \kappa_n$
with $\kappa_n>0$, and correspond to
the $N$ real eigenvalues $E_n=-\kappa_n^2$, $1\leq n\leq N$, with $-V_0<E_n<0$.
The other roots are located symmetrically in the negative half-plane
$\Im m(k)<0$, in such
a way that if $k$ is a root, $-\overline{k}$ is also a root of (\ref{trans}).

Numerically, the pure imaginary roots $k_n$ can be computed by using
Newton's method.
For the other complex roots, one can show that~\cite{nus} the behaviour of
$\xi_n=k_nr_0=x_n+i y_n$ for $n$ large, is given by:
\begin{equation}
{\displaystyle \left\{ \begin{array}{l}
  \ds x_n=c_n-\frac{1}{c_n}\log (\frac{2c_n}{A}+...   ,\\[4mm]
  \ds y_n=-\log (\frac{2c_n}{A})-\frac{1}{2c_n^2}
                      (\log (\frac{2c_n}{A})-1)^2+...  \\
                      \end{array}\right.}
\label{asym}
\end{equation}
where $c_n=(n+\frac{1}{2})\pi$.
Then $\xi_n \sim n\pi-i \log n$ gives an initial guess for the large
order roots.

\section{Matrix elements of the potentials:}

\subsection*{Gaussian Potential}

To express the matrix elements of the Gaussian potential in compact form,
we introduce the function:
\begin{equation}
K_{\alpha\beta}^L(\gamma)=-\frac{\exp(\beta\gamma+\frac{\gamma^2}{4\alpha})
  \sqrt{\pi}}{{8\sqrt{\alpha}}}
  \biggl[\erf\Bigl(\sqrt{\alpha}(r-\beta-\frac{\gamma}{2\alpha})\Bigr)
  +\erf\Bigl(\sqrt{\alpha}(\beta+\frac{\gamma}{2\alpha})\Bigr),
  \biggr]
\end{equation}
where erf denotes the error function~\cite{abst}. The matrix elements of
a general gaussian function:
\begin{equation}
f_{\alpha\beta}(r)=e^{-\alpha(r-\beta)^2},
\end{equation}
can be written:
\begin{align}
\begin{split}
  \ds\int_0^L\sin(\kappa_i^*r)f_{\alpha\beta}(r)\sin(\kappa_j r)\,&dr=
            K_{\alpha\beta}^L\Bigl(i(\kappa_i^*+\kappa_j)\Bigr)
           +K_{\alpha\beta}^L\Bigl(-i(\kappa_i^*+\kappa_j)\Bigr) \\[4mm]
   &\ \ \ \ \ds -K_{\alpha\beta}^L\Bigl(i(\kappa_i^*-\kappa_j)\Bigr)
            -K_{\alpha\beta}^L\Bigl(-i(\kappa_i^*-\kappa_j)\Bigr). \\
\end{split}
\end{align}
If we choose $(\alpha=0.25,\beta=3.5)$ and $(\alpha=0.2,\beta=0)$,
we can use this relation to compute the matrix element of the potential
(\ref{neqpara}) very simply.

\subsection*{Saxon-Woods potential}

In order to compute the matrix elements of the Saxon-Woods
potential, we introduce the function $I_{r_0a}^{L}(\kappa)$
defined by:

\begin{equation}
I_{r_0a}^{L}(\kappa)
=\int_0^L\frac{e^{i\kappa r}}{1+e^\frac{r-r_0}{a}}dr.
\end{equation}

This integral can be evaluated using the hypergeometric
function $_2F_1$~\cite{abst}:
\begin{align}
  \begin{split}
    I_{r_0a}^{L}(\kappa)
    =\frac{i}{\kappa}\biggl[
    \,_2F_1\bigl(1,ia\kappa;1+ia\kappa;-&e^{-\frac{r_0}{a}}\bigr) \\[1mm]
    -&e^{i\kappa L}\,_2F_1\bigl(1,ia\kappa;1
      +ia\kappa;-e^{\frac{L-r_0}{a}}\bigr)
    \biggr].
  \end{split}
\label{fcti}
\end{align}

\g The matrix element of a Saxon-Woods potential $V(r)$ with depth
$V_0$, diffuseness $a$ and radius $r_0$ are given by:

\begin{equation}
\mathbf V_{mn}=-V_0\int_0^L\frac{\sin(\kappa_m^*r)\sin(\kappa_n r)}{1+e^\frac{r-r_0}{a}}dr,
\end{equation}

\g and can be written explicitly using the $I_{r_0a}^{L}(\kappa)$
functions:

\begin{equation}
\mathbf V_{mn}=-\frac{V_0}{2}\Re\biggl[I_{r_0a}^{L}(\kappa_m^*-\kappa_n)-I_{r_0a}^{L}(\kappa_m^*+\kappa_n)\biggr].
\end{equation}

\g If $\kappa_m^*=\kappa_n$, this expression becomes:

\begin{equation}
\mathbf V_{mn}=-\frac{V_0}{2}\biggl[L+a\ln\Bigl[\frac{1+e^\frac{r_0}{a}}{e^\frac{L}{a}+e^\frac{r_0}{a}}\Bigr]
-\frac{I_{r_0a}^{L}(2\kappa_n)+I_{r_0a}^{L}(-2\kappa_n)}{2}\biggr].
\end{equation}

\subsection*{Derivative of Saxon-Woods potential}

We have used for the particle-particle channel a potential
with the shape of the derivative of a Saxon-Woods potential
with depth $\Delta_0$, diffuseness $a$ and radius $r_0$:
\begin{equation}
\Delta(r)=\Delta_0\biggl[\frac{d}{dr}\frac{1}{1+e^\frac{r-r_0}{a}}\biggr].
\end{equation}

The matrix elements of this potential can also be written using
the function~(\ref{fcti}):
\begin{align}
\begin{split}
  \mathbf\Delta_{mn}
   & =\Delta_0\Biggl[\frac{\kappa_m^*-\kappa_n}{4i}
     \Bigl[I_{r_0a}^{L}(\kappa_m^*-\kappa_n)
          -I_{r_0a}^{L}(\kappa_n-\kappa_m^*)\Bigr] \hskip 1cm \\
\noalign{\vskip 2mm}
   &\hskip 0.05cm
+\frac{\sin(\kappa_m^*L)\sin(\kappa_nL)}{1+\exp(\frac{L-r_0}{a})}
    -\frac{\kappa_m^*+\kappa_n}{4i}
     \Bigl[I_{r_0a}^{L}(\kappa_m^*+\kappa_n)
          -I_{r_0a}^{L}(-\kappa_m^*-\kappa_n)\Bigr]
   \Biggr]
\end{split}
\end{align}

\subsection*{Centrifugal term}

The matrix elements associated with the centrifugal term are
given by:

\begin{equation}
\mathbf C_{ij}=\frac{\hbar^2\ell(\ell+1)}{2m}\int_0^L\frac{\sin\kappa_i^*r\sin\kappa_j r}{r^2}dr.
\end{equation}

\g This integral can be evaluated using the sine integral
function~\cite{abst}:
\begin{align}
  \begin{split}
\mathbf C_{ij}=-\frac{\hbar^2\ell(\ell+1)}{2m}
\biggl[
\frac{\sin\kappa_i^*L\sin\kappa_jL}{L}
 + &\frac{\kappa_i^*-\kappa_j}{2}\hbox{Si}(\kappa_i^*-\kappa_j)L \\
&\quad - \frac{\kappa_i^*+\kappa_j}{2}\hbox{Si}(\kappa_i^*+\kappa_j)L
\biggr].
  \end{split}
\end{align}

\end{document}